\documentclass[aps,pre,twocolumn,numerical,groupedaddress]{revtex4-1}
\usepackage{hyperref,lineno,natbib,amsmath,amssymb,graphicx,color,microtype}
\usepackage{mathptmx}

\usepackage{hyperref,color}

\definecolor{webgreen}{rgb}{0,.35,0}
\definecolor{webbrown}{rgb}{.6,0,0}
\definecolor{RoyalBlue}{rgb}{0,0,0.9}
\definecolor{mywhite}{rgb}{1.0,1.0,1.0}
\newcommand{\white}[1]{\textcolor{mywhite}{#1}}

\hypersetup{
   colorlinks=true, linktocpage=true, pdfstartpage=3, pdfstartview=FitV,
   breaklinks=true, pdfpagemode=UseNone, pageanchor=true, pdfpagemode=UseOutlines,
   plainpages=false, bookmarksnumbered, bookmarksopen=true, bookmarksopenlevel=1,
   hypertexnames=true, pdfhighlight=/O,
   urlcolor=webbrown, linkcolor=RoyalBlue, citecolor=webgreen,
   pdfauthor={Chris H. Rycroft and Martin Z. Bazant},
   pdftitle={Asymmetric collapse by dissolution or melting in a uniform flow},
   pdfkeywords={},
   pdfcreator={pdfLaTeX},
   pdfproducer={LaTeX with hyperref}
}

\definecolor{dblue}{rgb}{0,0,0.8}
\definecolor{dgreen}{rgb}{0,0.55,0}
\definecolor{dred}{rgb}{0.8,0,0}
\newcommand{\dblue}[1]{\textcolor{dblue}{#1}}
\newcommand{\dgreen}[1]{\textcolor{dgreen}{#1}}
\newcommand{\dred}[1]{\textcolor{dred}{#1}}

\definecolor{dpurp}{rgb}{0.4,0,0.7}

\renewcommand{\vec}[1]{\mathbf{#1}}
\newcommand{\ob}[1]{\overline{#1}}
\newcommand{\ep}{e^{i\theta}}
\newcommand{\en}{e^{-i\theta}}
\newcommand{\zb}{\bar{z}}
\newcommand{\wb}{\bar{w}}
\newcommand{\Pb}{\bar{P}}
\newcommand{\N}{\mathbb{N}}
\newcommand{\oow}{\left(\tfrac{1}{w}\right)}
\newcommand{\inttp}{\int_0^{2\pi}}
\newcommand{\Pe}{\text{Pe}}
\newcommand{\Peh}{\widetilde{\Pe}}
\newcommand{\Rey}{\text{Re}}
\newcommand{\Sch}{\text{Sc}}
\newcommand{\Pra}{\text{Pr}}
\newcommand{\nor}{\hat{\vec{n}}}
\newcommand{\vx}{\vec{x}}
\newcommand{\vv}{\vec{v}}
\newcommand{\bq}{\bar{q}}
\newcommand{\ur}{u_\text{r}}
\newcommand{\pheq}{\phantom{={}}}
\DeclareMathOperator{\Real}{Re}
\DeclareMathOperator{\Imag}{Im}

\begin{document}
\title{Asymmetric collapse by dissolution or melting in a uniform flow}
\author{Chris H. Rycroft}
\email{chr@seas.harvard.edu}
\homepage{http://seas.harvard.edu/~chr/}
\affiliation{Paulson School of Engineering and Applied Sciences, Harvard University, Cambridge, MA 02138}
\affiliation{Department of Mathematics, Lawrence Berkeley Laboratory, Berkeley, CA 94720}
\author{Martin Z. Bazant}
\email{bazant@mit.edu}
\homepage{http://www.mit.edu/~bazant/}
\affiliation{Department of Chemical Engineering, Massachusetts Institute of Technology, MA 02139}
\affiliation{Department of Mathematics, Massachusetts Institute of Technology, MA 02139}
\date{\today}

\begin{abstract}
An advection--diffusion-limited dissolution model of an object being eroded by
a two-dimensional potential flow is presented. By taking advantage of the conformal
invariance of the model, a numerical method is introduced that tracks the
evolution of the object boundary in terms of a time-dependent Laurent series.
Simulations of a variety of dissolving objects are shown, which shrink and then
collapse to a single point in finite time. The simulations reveal a surprising
exact relationship whereby the collapse point is the root of a non-analytic
function given in terms of the flow velocity and the Laurent series
coefficients describing the initial shape. This result is subsequently derived
using residue calculus. The structure of the non-analytic function is examined
for three different test cases, and a practical approach to determine the
collapse point using a generalized Newton--Raphson root-finding algorithm is
outlined. These examples also illustrate the possibility that the model breaks
down in finite time prior to complete collapse, due to a topological
singularity, as the dissolving boundary overlaps itself rather than breaking up
into multiple domains (analogous to droplet pinch-off in fluid mechanics). In
summary, the model raises fundamental mathematical questions about broken
symmetries in finite-time singularities of both continuous and stochastic
dynamical systems.
\end{abstract}
\maketitle

\section{Introduction}
Interfacial growth processes, such as alloy
solidification~\cite{burden74,wollkind84,sethian92,theillard15},
electrodeposition~\cite{wranglen60,brady84}, and crystal
formation~\cite{kessler84,granasy04}, are responsible for a wide variety of
complex natural patterns~\cite{langer80,ben-jacob90} that emerge due to
instabilities in the underlying equations for interface
motion~\cite{mullins63,ben-jacob83}. Often, continuum models of interfacial
growth exhibit finite-time singularities, whereby features of the interface,
such as curvature, diverge after finite time. The formation of such
singularities is indicative of a breakdown in the separation of scales between
the macroscopic variables in the continuum model and the microscopic variables
that have been ignored~\cite{brenner00,zeff00}. In fluid mechanics, the
presence of finite-time singularities has been extensively studied, and can
frequently provide physical insight~\cite{barenblatt_book}, such as
identifying the existence of universal scaling behavior in the pinch-off of a
column of fluid~\cite{eggers93,brenner96}.

A particularly good example of interfacial growth is diffusion-limited
aggregation (DLA)~\cite{witten81}, where a solid cluster of particles is grown
from a bath of diffusing particles starting from a single static seed particle.
Additional particles are introduced one by one far away from the cluster, carry
out random walks until they adhere to the cluster upon contact, causing it to
grow. Since a random walker is more likely to first meet an extremity of the
cluster than an interior region, the extremities grow preferentially, leading
to complex fractal clusters in discrete computer simulations of the
model~\cite{witten81,meakin83a,meakin84,paterson84}.

Growth processes related to diffusion-limited aggregation have also been
studied in the continuum limit, whereby the steady-state walker concentration
satisfies Laplace's equation outside the cluster, is zero on the cluster
boundary, and tends to a steady concentration far away from the cluster, and
the cluster boundary grows continuously and deterministically with its velocity
proportional to the normal gradient of the diffusing concentration field. This
limit is mathematically equivalent to the classical Saffman--Taylor
problem~\cite{saffman58} of viscous fingering in a Hele-Shaw cell without
surface tension~\cite{bensimon86,tanveer00}. This model is conformally
invariant, which simplifies the analysis and allows it to be studied in detail
in two dimensions using conformal mapping~\cite{bazant05}, as first formulated
in 1945 by Polubarinova-Kochina~\cite{polub45a,polub45b} and
Galin~\cite{galin45} for applications to oil recovery and water filtration in
porous media. Continuous diffusion-limited growth is notoriously unstable, and
perturbations in an object's boundary progressively sharpen, eventually leading
to the formation of cusps in finite
time~\cite{shraiman84,feigenbaum01,davidovitch00}. In viscous fingering, these
finite-time singularities are regularized by surface
tension~\cite{tanveer93,siegel96,tanveer00}, which leads instead to tip
branching instabilities and the formation of fractal fingering
patterns~\cite{bensimon86}. 

Hybrid discrete--continuous models have also been developed based on iterated
conformal maps~\cite{hastings98}, which take full advantage of conformal
invariance in two dimensions~\cite{bazant05}. The growing cluster is defined by
a chain of conformal maps that each add a small bump to the shape to represent
the aggregation of a single  particle, which opens new possibilities, such as
growing non-random, fractal clusters~\cite{davidovitch00}. By studying the
statistics of the stochastic conformal map, it can also be shown that the
average shape of random DLA clusters is similar~\cite{tang85}, but not
identical~\cite{somfai03,davidovitch05}, to the corresponding shape of
continuous, deterministic diffusion-limited growth. 
 
Both the stochastic and continuous growth models have been extended to allow
for any gradient-driven transport process in two dimensions~\cite{bazant03}, on
flat or curved surfaces~\cite{choi10}, which is made possible by a general
conformal invariance principle~\cite{bazant04b}. In the canonical case of
advection-diffusion-limited aggregation (ADLA) of random walkers in a fluid
potential flow, asymptotic approximations of the flux profile have been
studied~\cite{choi05b}, and the discrete and continuous cases have been
compared~\cite{davidovitch05}. These studies of  ADLA provide the motivation
for the present work.

In this paper, we consider when the sign of growth is switched in the ADLA
model, corresponding to dissolution or erosion: we start with a solid object,
and then random walkers in a flow annihilate small parts of it on contact. Even
in the absence of flow, this case has received much less investigation, since
it usually leads to stable dynamics~\cite{paterson84,tang85,meakin86,krug91}
and thus many of the complex patterns due to growth instabilities are no longer
manifest. However, this model opens up alternative questions for study. In a
previous paper~\cite{bazant06a}, several different conformally invariant
transport-limited dissolution models were introduced, including the erosion of
corrugations on an infinite surface, and the expansion of a cavity due to
dissolution. The paper also introduced the system of
advection--diffusion-limited dissolution (ADLD), whereby an object is dissolved
due to a concentration of random walkers in a fluid potential flow past the
object. The object is represented by a time-dependent conformal map from the
unit circle to the physical domain. By making use of previous asymptotic
results~\cite{choi05b}, an evolution equation for the conformal map is derived
for the regime of intermediate P\'eclet number. The mathematical model is
similar to the model of freezing and melting of dendrites in a hydrodynamic
flow considered by Kornev and coworkers~\cite{kornev94,cummings99,cummings99a},
where the random walker concentration field in our model is replaced by a
temperature field governed by an advection--diffusion equation. However, these
works use a different mathematical formulation, employing the Schwarz
function~\cite{davis74} to model the time-dependent boundary, and consider a
different model for interfacial motion.

The previous study of ADLD~\cite{bazant06a} was entirely analytical, and
thus only considered the simple shapes of a circle and ellipse. Here, we
investigate this model in more depth, and develop a numerical implementation
that can simulate the dissolution of arbitrarily shaped objects. By simulating
arbitrary objects, we are able to investigate the model in substantially more
mathematical detail, particularly in relation to the formation of several
different types of finite-time singularity.

\subsection{Physical applications of the model}
While the model that we consider is for a mathematically idealized
two-dimensional case, it is worth considering what situations it could be
applied to. For free gas or liquid flow, the model is unlikely to apply.
Laminar flow at high Reynolds number $\Rey$ can be approximated as a potential
flow with viscous boundary layer of width \smash{$1/\sqrt{\Rey}$}, but our
model assumes that the diffusion layer of width \smash{$1/\sqrt{\Pe}$}, which
is much wider than any viscous boundary layer for the regime of intermediate
P\'eclet number that we consider. For melting, this would imply a small Prandtl
number $\Pra$, while for dissolution this would imply a small Schmidt number
$\Sch=\Pe/\Rey$. However, for most liquids $\Sch \gg \Pra \gg 1$, since
momentum diffuses much more rapidly than heat or mass. For gases, $\Sch \sim
\Pra \sim 1$ since the same collisional mechanism governs mass, momentum, and
heat transfer, but this still violates the model assumptions since the viscous
and diffusion boundary layers would have similar size.

One area where the model may apply is for dissolution or two-phase flow in
porous media, which has relevance to water transport in soils or to flow in oil
reservoirs, as in the seminal papers~\cite{polub45a,polub45b,galin45}. In this
case, the fluid can be modeled using Darcy flow, and the object would represent
a solidified region within the porous medium undergoing dissolution or melting
(as considered by Kornev and coworkers~\cite{kornev94,cummings99,cummings99a}).
A similar  model of nonlinear advection-diffusion in a potential flow is also
applicable to viscous gravity currents in Hele-Shaw cells or porous-media
flows, in which a heavier fluid spreads diffusively by gravity on an
impermeable surface, as it is sheared by the flow of a lighter fluid above
it~\cite{eames05}. The model could also have applications to a variety of
different electrochemical corrosion
processes~\cite{leger99,vukmirovic02,saunier04}, whenever a flow is imposed to
modify the diffusive transport of active ionic species.

The model may also be relevant to dissolution driven by electrokinetic
phenomena, in the regime where double layers are thin in comparison to the
object size. Consider a fixed object that is uniformly charged, in a fluid that
is driven by a uniform electric field. In this case the fluid motion will be
well-modeled by a potential flow where the electric potential is proportional
to the fluid potential. The model could also apply to an object moving via
electrophoresis. An object of constant surface charge will move at a constant
speed in a uniform electric field, regardless of its
shape~\cite{morrison-jr.70}, and hence constant flow at infinity could be
fictitious flow of a stagnant fluid in a frame of reference of a particle
moving at constant velocity by electrophoresis as it dissolves.

The model we consider is in a substantially different regime than recent
experiments and simulations of the erosion of clay bodies in high Reynolds
number fluid flows~\cite{ristroph12}, where the erosion rate of the
surface is proportional to shear stress, and the dissolving bodies tend towards
a self-similar cone-like shape~\cite{moore13}.

\subsection{Layout of the paper}
The paper proceeds as follows. In Section~\ref{sec:theory}, we present the
theoretical background of the model, and derive an evolution equation for the
shape of a dissolving body in terms of a time-dependent conformal map from the
unit circle to the physical domain, described by a Laurent series. In
Section~\ref{sec:numerics}, we then derive a system of ordinary differential
equations that govern how the Laurent series coefficients evolve with time. We
numerically integrate this system using an eighth-order timestepping
method~\cite{hairer_book,hairer_book2}, which allows the dissolution process to
be simulated very accurately, close to the limit of machine precision.

Our initial numerical results for a variety of objects show that they
completely dissolve in a finite duration with their boundaries becoming
progressively smoother (Section~\ref{sec:results}). As expected, the flow
causes the objects to dissolve more quickly on the side facing upstream,
although the details of the process are complicated, and affected by the
precise manner that the fluid flows past the dissolving object. Of particular
interest is the location of the collapse point, where the dissolving object
finally vanishes. Due to the high accuracy of our simulations, we inferred an
exact relationship between the collapse point $z_c$ expressed as a complex
number, the speed of the flow, and the initial Laurent coefficients. The
relationship is surprising, whereby $z_c$ is the root of a non-analytic
function $P$, the terms of which involve complicated products of Laurent series
terms. While some of these terms share similarities with binomial and
multinomial expansions, they are distinctly different, and we are unaware of
any other problem in conformal mapping or elsewhere where they occur.

In Section~\ref{sec:deriv} we make use of residue calculus to derive the
general form of $P$, using the numerical results as a guide. The complicated
products of terms in $P$ arise from the residue of a contour integral where
several Laurent series are multiplied together. In general, the function $P$
has multiple roots, thus creating ambiguity about which root is the collapse
point, and in Section~\ref{sec:examples} we consider three different example
objects that highlight the structure of $P$ in more detail. To find the roots
of $P$, we introduce a generalized Newton--Raphson iteration. As usual for
Newton--Raphson iterations, plots of the root convergence in terms of the
initial starting guess are fractal, but the non-analyticity of $P$ creates some
distinct morphological differences, and the plots illustrate the difficulties
of determining the collapse point with mathematical certainty.

The three examples also exhibit several types of finite-time singularity,
which are both physically relevant and provide insight into the mathematical
well-posedness of the model. Connections between these singularities and $P$
are discussed. While the dissolution model that we consider is a simplified
model with stable dynamics, it has a surprising amount of mathematical
structure, and our results raise a number of questions for further study.

\section{Theoretical background}
\label{sec:theory}
We make use of non-dimensionalized units, and consider an object in two
dimensions with a time-dependent boundary $S(t)$ as shown in
Fig.~\ref{fig:schem}(a). The object is immersed in an inviscid, irrotational
fluid with velocity $\vv(\vx,t)$, which can be written in terms of a potential
$\phi(\vx,t)$ as $\vv=\nabla \phi$. The fluid is incompressible, so $\nabla
\cdot \vv=0$ and hence
\begin{equation}
  \nabla^2 \phi =0.
  \label{eq:confeq1}
\end{equation}
At the boundary of the object the condition $\nor \cdot \vv = \nor \cdot \nabla
\phi =0$ is used, where $\nor$ is an outward-pointing normal vector. Far away
from the object the flow tends to a constant horizontal velocity so that
$\vv(\vx,t) \to (1,0)$ as $|\vx| \to \infty$. Equivalently, the potential 
satisfies $\phi(\vx,t) \to x$ as $|\vx| \to \infty$.

The fluid transports a random walker concentration $c(\vx,t)$ that satisfies
the advection--diffusion equation
\begin{equation}
  \Pe \nabla c \cdot \nabla \phi = \nabla^2 c,
  \label{eq:confeq2}
\end{equation}
where $\Pe$ is the P\'eclet number, a dimensionless quantity describing the
ratio of advection to diffusion. Far away from the object, the random walker
concentration tends to unity, so that $c(\vx,t) \to 1$ as $|\vx| \to \infty$.
The random walkers are responsible for dissolving the object. At the boundary
of the object, $c(\vx,0)=0$. The normal velocity of the object boundary $S(t)$
is given by
\begin{equation}
  \sigma = - \lambda \nor \cdot \nabla c,
  \label{eq:ad_bc}
\end{equation}
where $\lambda$ is a dimensionless constant. Equations~\eqref{eq:confeq1} and
\eqref{eq:confeq2} together with the associated boundary conditions form a
closed system for $(\phi,c,S)$ that describe the dissolution dynamics, but they
are difficult to solve directly. To proceed, we therefore treat the object as
being in the complex $z$ plane, where $z=x+iy$, and we introduce a
time-dependent conformal map described by an analytic function $z=g(w,t)$ that
transforms the unit circle $C$ into the object boundary $S(t)$, as shown in
Figure~\ref{fig:schem}(b). The most general form of the conformal map is the
truncated Laurent series,
\begin{equation}
  \label{eq:g}
  g(w,t) = a(t)w + \sum_{n=0}^N q_n(t) w^{-n}, 
\end{equation}
where $a(t)$ is taken to be a real function, and $q_n(t)$ are complex
functions. Hereafter, we refer to $q_n$ as the $n$th mode. Both
Eqs.~\eqref{eq:confeq1} and \eqref{eq:confeq2} are conformally invariant. The
Laplacian is the standard example of a conformally invariant operator, and the
advective term $\nabla c \cdot \nabla \phi$ is also conformally
invariant~\cite{bazant03,bazant04b}.

The boundary conditions in the $w$ plane are different. Due to the scaling
factor $a(t)$ in Eq.~\eqref{eq:g}, the boundary condition on the velocity
potential becomes $\phi(w,t) \to a \Real (w)$ as $|w| \to \infty$. We therefore
introduce a rescaled potential $\hat{\phi}(w,t) = \phi(w,t)/a$ that satisfies
the original boundary condition $\hat{\phi}(w,t) \to \Real (w)$ as $|w| \to
\infty$. The rescaled system for $c$ and $\hat{\phi}$ satisfies
Eqs.~\eqref{eq:confeq1} \& \eqref{eq:confeq2}, but with a rescaled P\'eclet number
\smash{$\Peh(t) = \Pe\, a(t)$}. In addition, the normal growth in the $w$ plane
is $\sigma_w = \sigma/|g'|$ to take into account the local volumetric scaling
of the conformal map.

Even in the $w$ plane where the object is the unit circle, the concentration
$c$ cannot be determined analytically. However, asymptotic expansions have been
studied in detail~\cite{choi05b}, and for P\'eclet numbers below $0.1$, the
approximation
\begin{equation}
  \sigma_w \sim \frac{\lambda I_0(\Peh) e^{\Peh \, \cos \theta}}{K_0\left(\frac{\Peh}{2}\right)} - \lambda \Peh \left( \cos \theta + \int_0^{\Peh} \frac{I_1(t)e^{t\cos \theta}}{t} dt \right),
\end{equation}
is uniformly accurate in $\theta = \arg w$. Taking the leading term of this
approximation gives
\begin{equation}
  \sigma_w \sim \frac{\lambda (1+ \Peh \, \cos \theta)}{-\gamma -\log \frac{\Peh}{4}} - \lambda \Peh \,\cos \theta,
  \label{eq:approx_lead}
\end{equation}
where $\gamma$ is Euler's constant.

To make progress, we now focus on the intermediate regime starting at small
P\'eclet number and ending prior to collapse, in which it is reasonable to
assume that \smash{$\log \Peh$} is a constant~\cite{bazant06a}. By rescaling
the time, we choose \smash{$\lambda= \gamma+ \log \frac{\Peh}{4}$} without loss
of generality. If the constant $B=\Pe \lambda$ is introduced, which we
subsequently refer to as the flow strength, then Eq.~\eqref{eq:approx_lead}
becomes
\begin{equation}
  \label{eq:sigw}
\sigma_w = -1 + Ba(t) \cos \theta.
\end{equation}
To transform this back into the physical domain, consider a point on the
$z(t)=g(w(t),t)$ on the boundary $S(t)$ of the object. Taking a time derivative
gives $\dot{z}=g'\dot{w}+\dot{g}$. Multiplying by \smash{$\overline{wg'}$} and
taking the real part gives
\begin{equation}
  \label{eq:pg2}
  \Real(\overline{w g'} \dot{z}) = \Real (\overline{w g'} g' \dot{w}) + \Real(\overline{w g'} \dot{g}).
\end{equation}
Since the point in the $w$ plane mapping to $z(t)$ must lie on the unit circle,
it follows that $w\wb=1$ and hence $\Real(\wb \dot{w})=0$, so the first term
on the right hand side of Eq.~\eqref{eq:pg2} vanishes. The motion of the point in
the $z$ plane is $\dot{z} = \sigma \hat{n}$ where $\hat{n}$ is the normal
vector written as a complex number. Taking into account rotation and scaling,
the normal vector is given by
\begin{equation}
\hat{n} = \frac{g'}{|g'|} \,\frac{w}{|w|}
\end{equation}
and hence the left hand side of Eq.~\eqref{eq:pg2} is
\begin{equation}
  \Real (\overline{wg'} \dot{z}) = \Real\left(\frac{\overline{wg'} g'w \sigma}{|g'w|} \right) = \Real( |g'| \sigma) = \sigma_w.
  \label{eq:pg3}
\end{equation}
Combining Eqs.~\eqref{eq:sigw}, \eqref{eq:pg2}, and \eqref{eq:pg3} yields 
\begin{equation}
  \label{eq:time_ev}
  \Real(\overline{wg'} \dot{g}) = -1 + B a(t) \cos \theta,
\end{equation}
which describes the dissolution process in terms of a time-dependent conformal
map. For $B=0$ it becomes the Polubarinova--Galin
equation~\cite{polub45a,polub45b,galin45}, which has been used in previous
continuum DLA (viscous fingering) studies without
advection~\cite{shraiman84,bensimon86}. The incorporation of the $Ba(t) \cos
\theta$ term~\cite{bazant06a} represents the simplest extension to account for
the general effect of advection~\cite{bazant03} and is therefore a useful and
interesting model to study in its own right.

\setlength{\unitlength}{0.88bp}
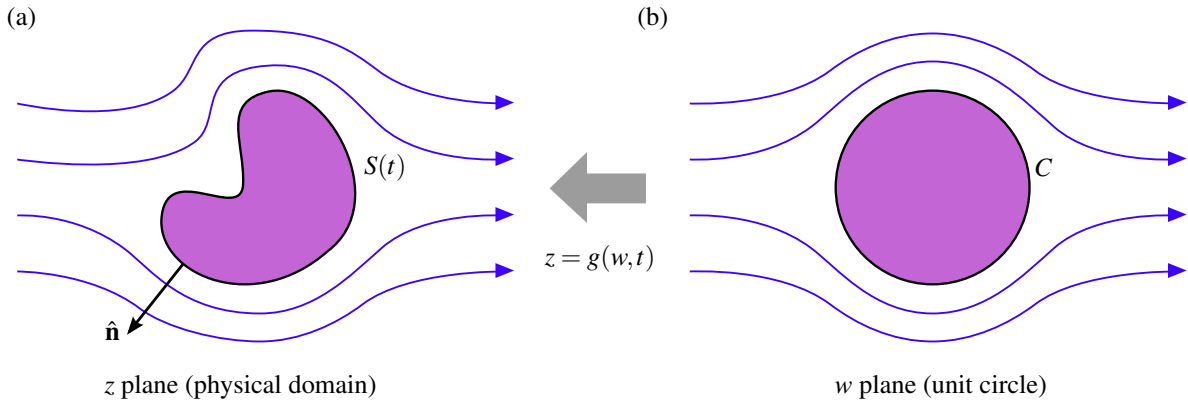
\begin{figure*}
  \begin{center}
    {\normalsize
    \begin{picture}(510,160)(0,0)
      \put(0,20){\includegraphics[scale=0.88]{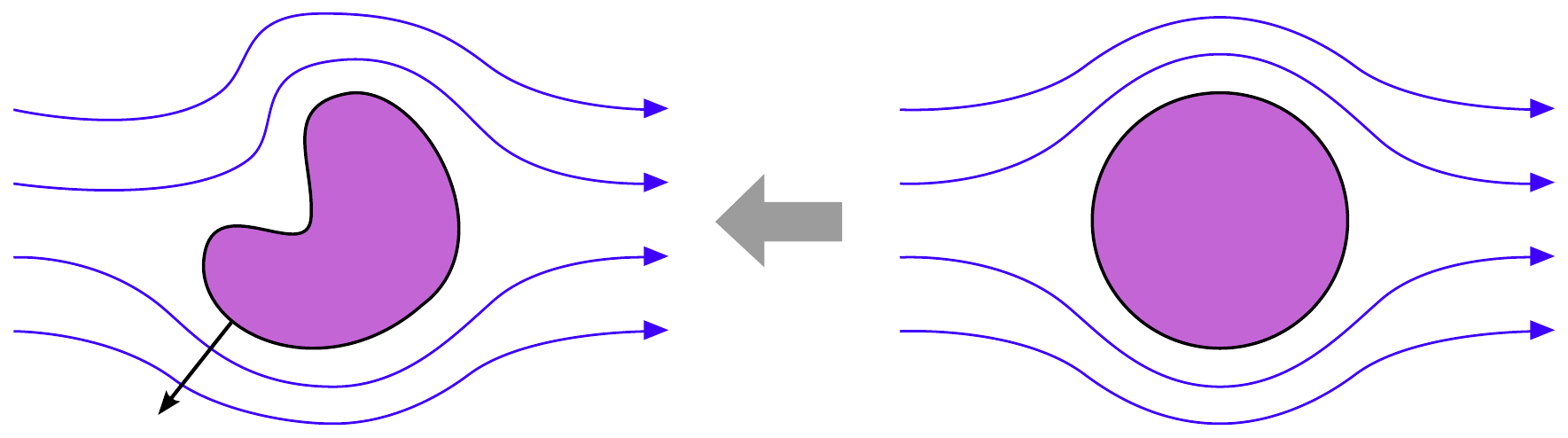}}
      \put(100,5){\makebox(0,0)[c]{$z$ plane (physical domain)}}
      \put(400,5){\makebox(0,0)[c]{$w$ plane (unit circle)}}
      \put(162,99){\makebox(0,0)[c]{$S(t)$}}
      \put(444,99){\makebox(0,0)[c]{$C$}}
      \put(254,60){\makebox(0,0)[c]{$z=g(w,t)$}}
      \put(0,160){(a)}
      \put(270,160){(b)}
      \put(42,25){$\nor$}
    \end{picture}}
  \end{center} \vspace{-4mm}
  \vspace{4pt}
  \caption{(a) The physical problem considered, where a two-dimensional object
  with time-dependent boundary $S(t)$ is dissolved a chemical concentration
  being transported by an incompressible potential flow. (b) A reference domain
  of the same physical problem but where the boundary is the unit circle $C$. A
  time-dependent conformal map $z=g(w,t)$ describes the transformation between
  the two domains.\label{fig:schem}}
\end{figure*}

\section{Numerical method and implementation}
\label{sec:numerics}
\subsection{Discrete formulation of the governing equation}
We now make use of Eq.~\eqref{eq:time_ev} to formulate a numerical solution
technique. We represent the dissolving object via the time-dependent conformal
map in Eq.~\eqref{eq:g} with a fixed value of $N\ge1$. We write the $q_n(t)$ in
component form as $b_n(t)+ic_n(t)$, and describe the shape of the object by the
real vector $\vec{s}(t)=(a,b_0,b_1,\ldots,b_N, c_0,c_1,\ldots,c_N)$, with a
total of $2N+3$ components. Using the two expressions
\begin{align}
\overline{wg'} &= a \bar{w} - \sum_{n=0}^N n(b_n-ic_n) \bar{w}^{-n}, \qquad \\
\dot{g} &= \dot{a} + \sum_{n=0}^N (\dot{b}_n + i \dot{c}_n) w^{-n},
\end{align}
Eq.~\eqref{eq:time_ev} becomes
\begin{align}
  \label{eq:pgcomp}
  -1+Ba\cos\theta &= \Real\Bigg(\left[ a e^{-i\theta} - \sum_{n=0}^N n
  (b_n-ic_n) e^{in\theta}\right] \nonumber \\
  &\pheq \left[\dot{a} e^{i\theta} + \sum_{n=0}^N
  (\dot{b}_n + i \dot{c}_n) e^{-in\theta}\right]\Bigg).
\end{align}
Eq.~\eqref{eq:pgcomp} is real, and can be expressed in terms of components $\cos
n\theta$ and $\sin n\theta$ for $n=1,\ldots,N+1$, plus a constant term.
Equating both sides of the Eq.~\eqref{eq:pgcomp} in each component leads to
$2N+3$ coupled ordinary differential equations for the $2N+3$ variables $a(t)$,
$b_n(t)$, and $c_n(t)$. Hence, other than for cases where these equations
are degenerate, $\dot{\vec{s}}$ will be uniquely determined in terms of
$\vec{s}$. Furthermore, since Eq.~\eqref{eq:pgcomp} does not feature any
higher harmonic of sine and cosine, it follows that $\vec{s}$ exactly represents the
time-evolution prescribed by Eq.~\eqref{eq:time_ev}: if a shape initially is
described in terms of a Laurent series using terms up to $q_N$, it will remain
perfectly described by this Laurent series throughout the whole dissolution
process.

The details of equating each component of Eq.~\eqref{eq:pgcomp} are given in
Appendix
\ref{app:numerics}. Equating the constant terms gives 
\begin{equation}
  \label{eq:teva}
  a\dot{a} - \sum_{n=0}^N n(b_n\dot{b}_n + c_n \dot{c}_n) = -1.
\end{equation}
Equating the terms with factors $\cos(N+1) \theta$ and $\sin(N+1)\theta$
gives
\begin{equation}
  \label{eq:thighest}
a \dot{b}_N = \dot{a}Nb_N, \qquad a \dot{c}_N = \dot{a}Nc_N,
\end{equation}
respectively. Equating the terms with a factor of $\sin n\theta$ for
$n=1,\ldots,N$ gives
\begin{align}
  0&= -\dot{a} (n-1) c_{n-1} + a \dot{c}_{n-1} \nonumber \\
  &\pheq - \sum_{m=0}^{N-n} \Big[(m+n)(c_{m+n} \dot{b}_m - b_{m+n} \dot{c}_m) \nonumber \\
  &\pheq - m(c_m \dot{b}_{m+n} - b_m
  \dot{c}_{m+n})\Big]
  \label{eq:tevsin}
\end{align}
Finally, equating the terms with a factor of $\cos n\theta$ for $n=1,\ldots,N$
gives
\begin{align}
  \beta_n&=-\dot{a} (n-1) b_{n-1} + a \dot{b}_{n-1} \nonumber \\
  &\pheq- \sum_{m=0}^{N-n}
  \Big[(m+n)(b_{m+n}\dot{b}_m + c_{m+n} \dot{c}_m) \nonumber \\
  &\pheq + m (b_m \dot{b}_{m+n} + c_m
  \dot{c}_{m+n})\Big]
  \label{eq:tevcos}
\end{align}
where $\beta_n = Ba$ if $n=1$, and $\beta_n=0$ otherwise. The combination
of Eqs.~\eqref{eq:teva}, \eqref{eq:thighest}, \eqref{eq:tevsin}, and \eqref{eq:tevcos}
can then be expressed as a linear system
\begin{equation}
  \label{eq:mat}
  M(\vec{s}) \dot{\vec{s}} = \vec{v}(\vec{s})
\end{equation}
where $M$ and $\vec{v}$ are matrix and vector functions of $\vec{s}$,
respectively. By writing Eq.~\eqref{eq:mat} as $\dot{\vec{s}}= M^{-1}(\vec{s})
\vec{v}(\vec{s})$, the system can be integrated numerically.

\subsection{Numerical implementation}
The simulations are carried out using double-precision floating point
arithmetic, using \textsc{LAPACK}~\cite{anderson99} to invert the linear system
in Eq.~\eqref{eq:mat}. To time-integrate the equation, the \textsc{DOP853}
integration routine described by Hairer {\it et al.}~\cite{hairer_book} is
used. This routine uses the eighth-order, thirteen-step Dormand--Prince
integration method that has the first-same-as-last (FSAL) property, requiring
twelve function evaluations per timestep~\cite{dormand89}. As described in more
detail later, the components of $\vec{s}$ can sometimes vary rapidly,
particularly close to the time of collapse. The \textsc{DOP853} routine employs
adaptive timestepping, which can retain accuracy in this situation. The routine
estimates the local
error~\footnote{The local error is defined as
\smash{$(\sum_{i=1}^{2N+3} e_i^2)^{1/2}$} where $e_i$ is the estimated error of
the $i$th component of $\vec{s}$ during a single timestep.} using a combination
of fifth-order and third-order embedded numerical schemes. For all of the
subsequent results, the timestep size $\Delta t$ is continually adjusted so
that the absolute local error per timestep remains below a tolerance of
$10^{-14}$. If the estimated error of a timestep exceeds the tolerance, then
the timestep is rejected and the integrator tries again with a reduced $\Delta
t$.

There are three scenarios where the \textsc{DOP853} integrator terminates
early: (i) if a maximum number of timesteps is reached, (ii) if the equations
are detected as stiff~\cite{hairer_book2}, or (iii) if the timestep $\Delta t$
required achieve the desired local error becomes too small. In the following
results, we have only observed the third scenario. This occurs when $\Delta t$
becomes smaller than $10 \ur t$, where $\ur=2.3\times10^{-16}$ is an estimate
of the smallest number satisfying $1.0+\ur>1.0$ in double-precision floating
point arithmetic. In certain cases, such as the examples of
Subsecs.~\ref{sub:ex1} and \ref{sub:ex3}, the third scenario signifies a
breakdown of the physical problem due to the formation of a cusp. However, the
third scenario also occurs in normal cases close to the time of collapse
$t_c$ due to $a(t)$ varying rapidly. If the \textsc{DOP853} integrator
terminates within $10^4\ur$ of $t_c$ then we manually advance to $t_c$ using
timesteps of $10\ur$ or less. While this may no longer achieve the required
level of local error, we find that it provides several additional digits of
accuracy in the collapse point location, which is useful in some of the later
analysis.

In some of the subsequent results, we must evaluate $\vec{s}$ at time points
spaced at fixed intervals, which may not precisely coincide with the time
points that are selected during the adaptive time-integration, which are
usually unevenly spaced. To solve this we use the dense output formulae
described by Hairer {\it et al.}~\cite{hairer_book}. By performing three
additional integration steps, the solution can be approximated as a
seventh-order polynomial over the interval of a timestep, allowing $\vec{s}$ to
be evaluated at any specific time point. For computational efficiency, these
three additional steps are only done when one or more output time points
overlaps with the current timestep interval.

The simulations are implemented in C++, and the code required to perform all of
the subsequent analysis is provided as Supplementary Information. For all of
the results presented here, the computation time required to simulate the
dissolution process is negligible, taking less than 0.25~s on a Mac Pro (Late
2013) with an 8-core 3~GHz Intel Xeon E5 processor.

\section{Results}
\label{sec:results}
\subsection{Analytic results for the area and highest mode amplitude}
Before presenting results of the numerical method, it is useful to establish
some basic features of the equations presented in the previous section. The
area of the object is given by
\begin{equation}
  A(t) = \iint_\Omega dz
\end{equation}
where $\Omega$ is the region enclosed by $S(t)$. Using Green's identity in
complex form,
\begin{equation}
  \label{eq:carea}
  A(t) = -\frac{1}{2i} \oint_{S(t)} z d\zb = \frac{1}{2i} \int_C g(w) \overline{g'(w)} d\wb.
\end{equation}
Since $w\wb=1$ on the unit circle, the integrand can be converted into an
analytic function,
\begin{align}
  A(t) &= \frac{1}{2i} \int_C g(w) \overline{g'(w)} \frac{dw}{w^2} \nonumber \\
  &= \frac{1}{2i} \int_C \left( a - \sum_{n=0}^N q_n n w^{-(n+1)}\right )\left( \frac{a}{w} + \sum_{n=0}^N \bq_n w^{n} \right) dw
\end{align}
and applying residue calculus gives
\begin{equation}
  A(t) = \pi \left( a^2 - \sum_{n=0}^N n|q_n|^2 \right) = \pi \left( a^2 - \sum_{n=0}^N n(b_n^2+c_n^2) \right),
\end{equation}
describing the area as a function of the current mode amplitudes. Furthermore,
time-integrating Eq.~\eqref{eq:teva} gives
\begin{equation}
  \label{eq:ctime}
  a^2 - \sum_{n=0}^N n(b_n^2 + c_n^2) = C-2t,
\end{equation}
where $C$ is a constant, and hence
\begin{equation}
  A(t) = A_0 - 2\pi t,
\end{equation}
where $A_0$ is the initial area of the object. The area of the object therefore
decreases at a constant rate, independent of the flow parameter $B$, with the
time to collapse given by
\begin{equation}
  t_c = \frac{A_0}{2\pi} = \frac{1}{2} \left( a^2 - \sum_{n=0}^N n|q_n|^2 \right).
  \label{eq:tc}
\end{equation}
The modes in Eq.~\eqref{eq:thighest} also have first integrals,
\begin{equation}
  b_N = k a^N, \qquad c_N = l a^N
  \label{eq:thighest2}
\end{equation}
for some constants $k$ and $l$. The highest mode amplitudes are therefore only
dependent on the conformal radius $a$. Due to the couplings in
Eqs.~\eqref{eq:tevsin} and \eqref{eq:tevcos}, similar results for the lower modes
do not exist.

\subsection{Initial numerical results}
\begin{figure*}
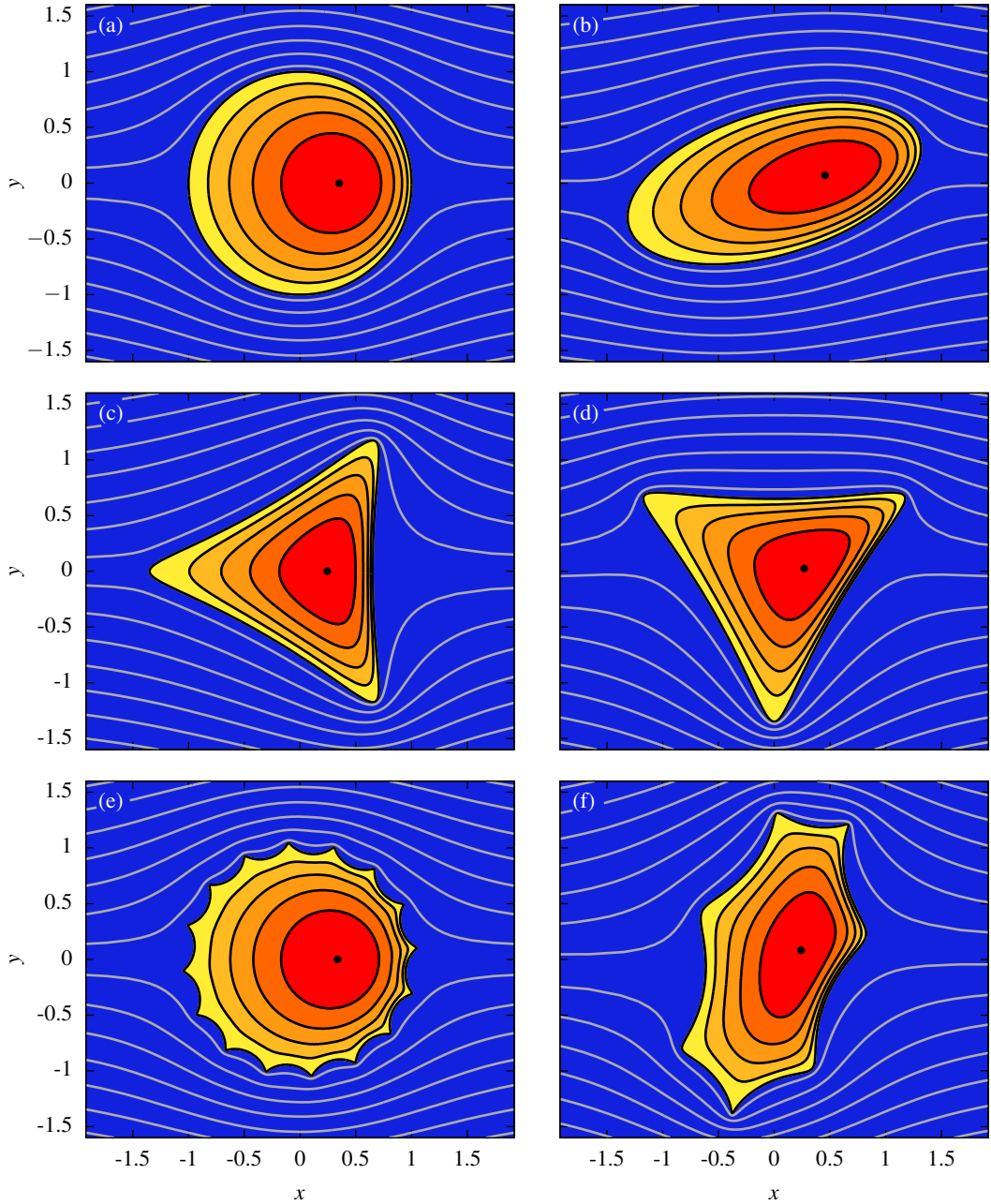

  \begin{center}
    {\small
    \include{basic_examples}}
  \end{center} \vspace{-4mm}
  \caption{Sample dissolution processes for six objects, all starting with
  $a=1$ and using $B=0.7$. The six objects and initial non-zero modes are (a) a
  circle, (b) an ellipse with $q_1=0.3+0.2i$, (c) a triangle with $q_2=-0.35$,
  (d) a triangle with $q_2=0.35i$, (e) a corrugated circle with $q_{15}=0.05i$,
  and (f) an irregular object with $q_1=-0.28+0.2i$ and $q_6=0.1$. The white
  lines show the flow streamlines around the initial shape. The colored regions
  shown the shapes of the object at successive times as it dissolves, where each
  progressive region represents the dissolution of 20\% of the object's
  initial area. The black circles indicate the final points of
  collapse.\label{fig:basic_examples}}
\end{figure*}

Figure~\ref{fig:basic_examples} shows the dissolution process for six objects
calculated using the numerical code, where for all cases $a(0)=1$ and $B=0.7$.
Figure~\ref{fig:basic_examples}(a) shows the dissolution process for a circle.
Throughout the process, the circle retains its shape although its center
progressively moves rightward due to the effect of the flow, which
preferentially dissolves the side of the circle facing upstream. A similar
behavior is visible in Fig.~\ref{fig:basic_examples}(b) for an ellipse, which
keeps its shape through the dissolution process, while the ellipse center moves
up and right. The results for Figs.~\ref{fig:basic_examples}(a) and
\ref{fig:basic_examples}(b) match those that were previously studied
analytically~\cite{bazant06a}. For the case of no flow where $B=0$ where the
system is mathematically equivalent to (time-reversed) Laplacian growth, and
bubble contraction in a porous medium, it is known that the ellipse is the most
generic self-similar shape~\cite{mccue03,king09,mccue11}.

Figure~\ref{fig:basic_examples}(c) shows the dissolution process for a
triangular-shaped object given by setting $q_2=-0.35$ initially. In general,
the mode $q_n$ is responsible for an $(n+1)$-fold perturbation of the boundary.
If all of the $q_n$ are initially real, the object is symmetric about the $x$
axis, and will remain symmetric throughout the dissolution process. For the
case shown, the point of the triangle that faces upstream is more rapidly
dissolved than the other two. Unlike the previous two examples that retain
their shape during dissolution, the triangle becomes progressively more rounded
at later times. Figure~\ref{fig:basic_examples}(d) shows the dissolution
process when the previous object is rotated by $90^\circ$, which is achieved by
setting $q_2=0.35i$. This object is initially symmetric about the $y$ axis, but
the flow causes this symmetry to be lost as time passes. The collapse point is
slightly up and right from the origin.

Figure~\ref{fig:basic_examples}(e) shows the dissolution process for the case
when $q_{15}=0.05i$ initially, which creates a 16-fold perturbation in the
boundary. After $20\%$ of the object has dissolved, this perturbation is almost
completely removed, with the object's shape approaching that of a circle. This
is expected from Eq.~\eqref{eq:thighest2}, which shows that the highest mode will
be proportional to $a^N$ and hence decay more rapidly for larger $N$. If several
modes are initially non-zero as in Fig.~\ref{fig:basic_examples}(f) an
irregular shape is formed, which behaves like a combination of the previous
examples, with sharp features in the boundary being rapidly removed.

\begin{figure*}
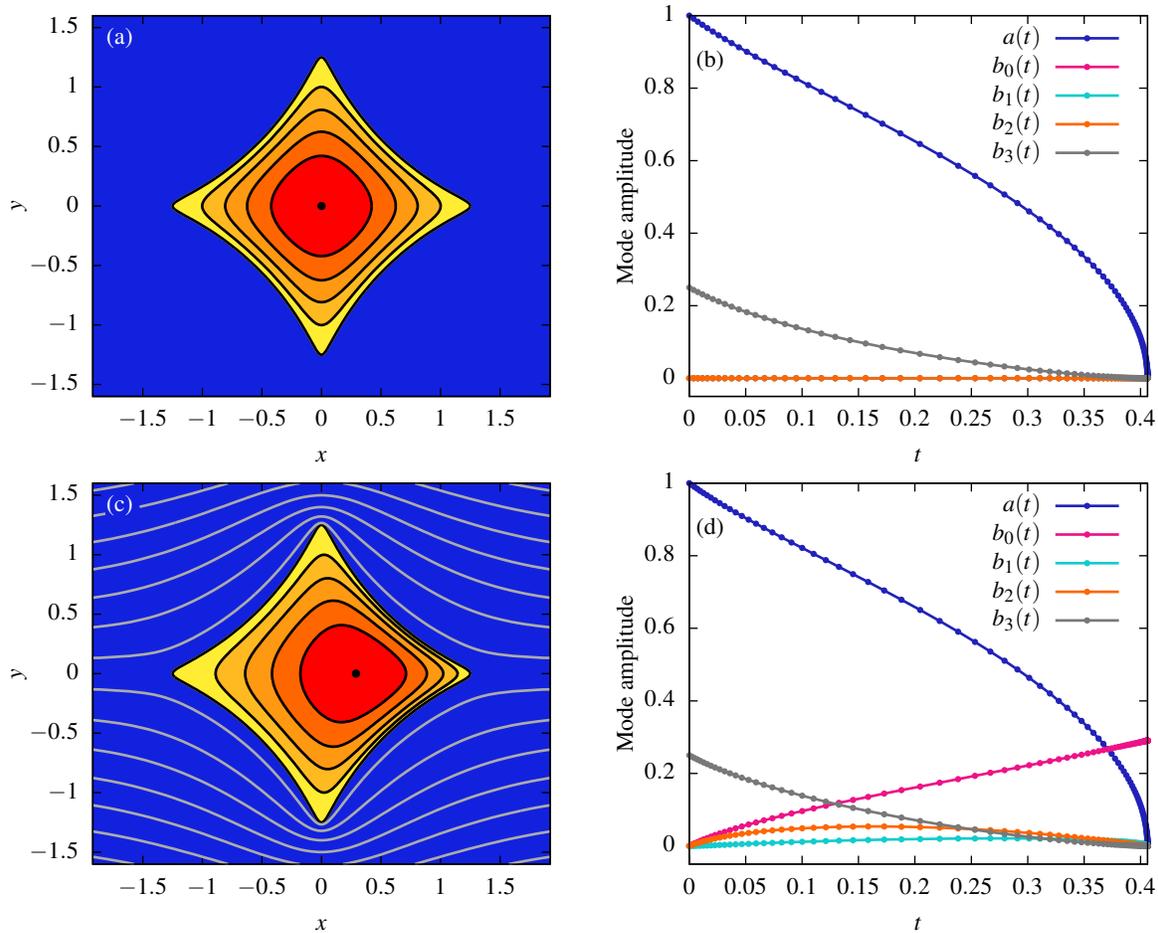

  \begin{center}
    {\small
    \include{mode_growth}}
  \end{center} \vspace{-4mm}
  \caption{(a) Dissolution process of a diamond shape initially described by
  non-zero modes $a=1$ and $q_3=0.25$ for the case of zero flow, $B=0$. The
  colored regions shown the shapes of the object at successive times as it
  dissolves, where each progressive region represents the dissolution of 20\%
  of the object's area, with the black circle indicating the point of collapse.
  (b) Time-evolution of the modes describing the diamond during dissolution,
  where the small circles on each curve show the integration timesteps using
  the adaptive \textsc{DOP853} integration scheme. (c) Dissolution process of
  the diamond when the flow is $B=0.7$. (d) Time-evolution of the modes
  describing the diamond during dissolution with flow, with the small circles
  on each curve showing the integration timesteps. \label{fig:mode_growth}}
\end{figure*}

We now examine the evolution of the modes and look in detail at the effect of
the flow strength $B$. We make use of the specific example of a diamond shape
given by $a=1$ and $q_3=0.25$ initially. Figure~\ref{fig:mode_growth}(a) shows
the dissolution process for the case of zero flow when $B=0$. Similar to
Figs.~\ref{fig:basic_examples}(c) and \ref{fig:basic_examples}(d) the object
becomes progressively more circular, but without the presence of flow it
retains symmetry in the $x$ axis, $y$ axis, and the line $x=y$.
Figure~\ref{fig:mode_growth}(b) shows the time-evolution of the modes
throughout the dissolution process. The modes $q_0$, $q_1$, $q_2$, which were
zero initially, remain zero throughout the dissolution process---this is
expected since any non-zero contribution from these modes would break at least
one of the symmetries seen in Fig.~\ref{fig:mode_growth}(a). The dissolution
process is therefore described entirely in terms of $a$ and $q_3$, and could
therefore be determined analytically using Eqs.~\eqref{eq:ctime} and
\eqref{eq:thighest2}, as considered in previous
work~\cite{shraiman84,howison86,bazant06a}. Since $q_0$ remains at zero, the
collapse point is at the origin.

Figure~\ref{fig:mode_growth}(c) shows the dissolution of the diamond when the
flow parameter is $B=0.7$. As in the previous examples of
Fig.~\ref{fig:basic_examples}, the diamond dissolves more rapidly on the side
facing upstream, and the collapse point is slightly downstream. The time
evolution of the modes (Fig.~\ref{fig:mode_growth}(d)) is significantly altered
in this case, with all three components $q_0$, $q_1$, and $q_2$ becoming
non-zero during the dissolution process, due to the mixing between modes via
the advection term in Eq.~\eqref{eq:time_ev}. The effects of these three modes,
such as the translation of the object center, and the loss of symmetry about
the $y$ axis, are clearly visible in Fig.~\ref{fig:mode_growth}(c). The $q_1$
and $q_2$ modes decay to zero at the point of collapse, while the $q_0$ mode
remains positive. The value of $q_0$ at $t=t_c$ gives the collapse point
position.

Figures~\ref{fig:mode_growth}(b) and \ref{fig:mode_growth}(d) also indicate the
adaptive integration timesteps chosen by the \textsc{DOP853} integration
routine. In the middle of the dissolution process, at $t\approx 0.2$, the
routine is able to take timesteps up to approximately 0.02 while retaining the
desired level of local error of $10^{-14}$. However, close to $t=t_c$, many
more timesteps are needed to resolve the rapid change in $a$. For the example
shown in Fig.~\ref{fig:mode_growth}(d), a total of 329 integration timesteps
are evaluated. During the \textsc{DOP853} integration routine, 194 steps are
accepted, and 128 are rejected due to the local error estimate exceeding the
given tolerance. Six additional small steps are required to reach the collapse
time $t_c$.

\begin{figure*}
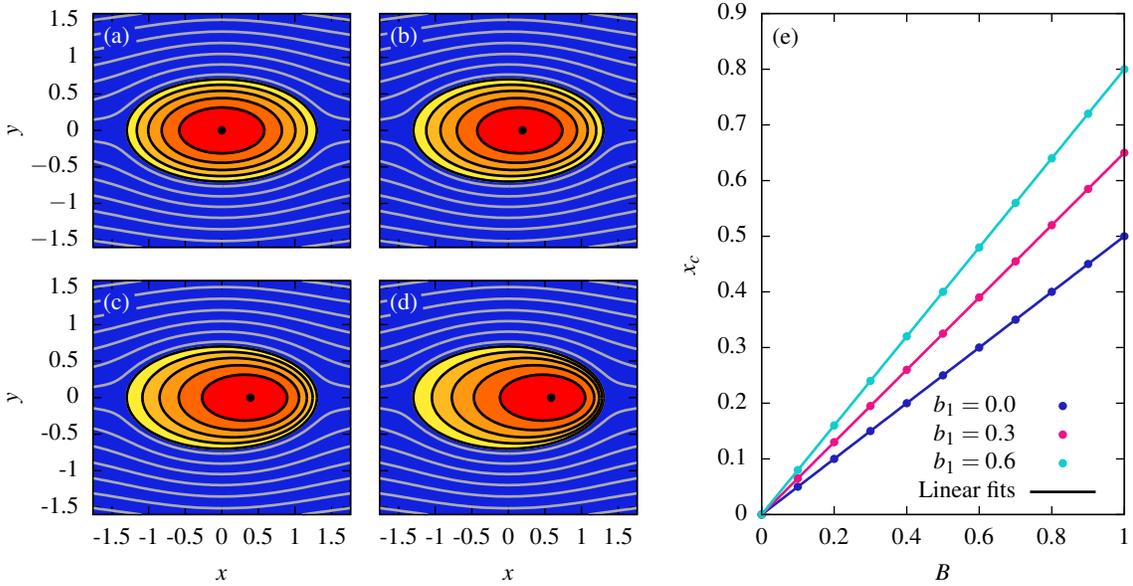

  \begin{center}
    {\small
    \include{fit_analytic}}
  \end{center} \vspace{-4mm}
  \caption{An example procedure to infer the analytic formula for position of
  collapse point in terms of the initial modes. (a--d) Dissolution processes of
  an ellipse given by $a=1, q_1=0.3$ for flow strengths of $B=0, 0.3, 0.6, 0.9$,
  respectively, using the same visual representation as described in
  Fig.~\ref{fig:basic_examples}. (e) Plot of the numerically computed horizontal
  collapse point position $x_c = \Real z_c$ as a function of $B$ for three
  different values of $b_1$, which match linear relationships to numerical
  precision, suggesting an analytical relationship.\label{fig:fit_analytic}}
\end{figure*}

\subsection{Inferring analytic formulae for the collapse point}
Figures~\ref{fig:basic_examples} and \ref{fig:mode_growth} show that the
collapse point $z_c=x_c+iy_c$ of the dissolution process is dependent on both
the flow strength $B$ and the initial shape of the body as described by its
Laurent coefficients. Since there are no other quantities in the problem, $z_c$
must be given in terms of $B$ and the Laurent coefficients only. The precise
form of this dependence is not obvious, as the collapse point is given as the
component $q_0$ of the nonlinear differential equation system, evaluated at the
time of collapse $t_c$.

In this section, we infer the exact form of this relationship by exploiting the
very high accuracy of the simulations, which allow the collapse point to be
calculated to at least twelve decimal places. To simplify the analysis, we set
$a=1$ throughout this section. To begin, we restrict to the case when the
Laurent coefficients are given purely in terms of real components $b_j$. As
discussed in the previous section, the components will remain real throughout
the simulation, and the object will be symmetric about the $x$ axis. Hence the
collapse point $z_c$ will be real, and determined entirely in terms of the
horizontal position $x_c$.

Figure~\ref{fig:fit_analytic} shows an example of inferring an analytic
relationship, for the case of an ellipse where the only non-zero Laurent
coefficient is $b_1$. Figures~\ref{fig:fit_analytic}(a--d) show four
dissolution process are shown for when $b_1=0.3$ and the flow strength is
$B=0,0.3,0.6,0.9$, respectively. The collapse point $x_c$ moves progressively
right as $B$ is increased. In Fig.~\ref{fig:fit_analytic}(e), the numerically
computed $x_c$ is plotted as a function of $B$, for three different values of
$b_1$ of $0.0$, $0.3$, and $0.6$. The plot demonstrates that $x_c$ is linear in
$B$, with the slope depending on $b_1$. The numerical data matches the 
relationship
\begin{equation}
  x_c=\frac{B(1+b_1)}{2},
  \label{eq:ell_infer}
\end{equation}
with the sum of square residuals being $1.3 \times 10^{-30}$, $1.3 \times
10^{-30}$, and $4.5 \times 10^{-30}$ for $b_1=0,0.3,0.6$, respectively. These
small residuals, which are of a similar size to the expected numerical error,
strongly suggest that this is an exact relationship for the original
mathematical problem.

One can extend this analysis to the case where the only non-zero Laurent
coefficient is $b_n$, and determine that $x_c$ satisfies the polynomial
relationship
\begin{equation}
  \frac{B}{2} = \frac{x_c-b_nx_c^n}{1-nb_n^2}.
\end{equation}
For $n=1$, this is consistent with Eq.~\eqref{eq:ell_infer} for the ellipse,
although it also reveals more structure, and by substituting Eq.~\eqref{eq:tc}
the relationship simplifies to 
\begin{equation}
  Bt_c = x_c-b_nx_c^n.
\end{equation}
To proceed, we now consider if there are two non-zero Laurent coefficients. The
simplest case would be $b_0$ and $b_n$ being non-zero, for $n\ge 1$. Since
$b_0$ corresponds to a translation, the relationship is immediately given by
\begin{equation}
  Bt_c = (x_c - b_0) - b_n (x_c-b_0)^n,
  \label{eq:zerof}
\end{equation}
without the need for simulation. Expanding the second term yields
\begin{align}
  Bt_c &= -b_n x_c^n + nb_n b_0 x_c^{n-1} - \tfrac{n(n-1)}{2} b_n b_0^2 x_c^{n-2} \nonumber \\
  &\phantom{={}} + \tfrac{n(n-1)(n-2)}{6} b_n b_0^3 x_c^{n-3} - \ldots + (x_c - b_0)
  \label{eq:pascal}
\end{align}
where the coefficients on the powers of $x_c$ follow Pascal's triangle.

The next case to consider is when $b_1$ and $b_n$ are non-zero, for $n \ge 2$.
Unlike the previous case this cannot be immediately derived, and must be
inferred through fitting to simulation. Table~\ref{tab:anal1} shows the derived
results for the cases of $n=2, 3, \ldots, 10$ where a surprising pattern
emerges. We see polynomials that bear some resemblance to a binomial expansion,
although in contrast to Eq.~\eqref{eq:pascal}, only every second power of $x_c$
is present. Furthermore, the coefficients in front of the terms are integer,
but of a more complicated form than Pascal's triangle. Unlike the previous
case, the more complicated form of these polynomials precludes rewriting them
in a succinct form like Eq.~\eqref{eq:zerof}. The pattern continues for the case
when $b_2$ and $b_n$ are non-zero, for $n \ge 3$. As shown in
Table~\ref{tab:anal2}, only every third power of $x_c$ is present. The integer
coefficients follow a natural progression from those in Table~\ref{tab:anal1}.

\begin{table*}
  \begin{center}
    \normalsize
    \begin{tabular}{l|l}
      $n$ & $Q(x_c)$ \\
      \hline
      2 & $-b_2x_c^2 + \dblue{2}b_2b_1 + (1-b_1) x_c$ \\
      3 & $-b_3x_c^3 + \dblue{3}b_3b_1 x_c + (1-b_1) x_c$ \\
      4 & $-b_4x_c^4 + \dblue{4}b_4b_1 x_c^2 - \dgreen{2}b_4b_1^2 + (1-b_1)x_c$  \\
      5 & $-b_5x_c^5 + \dblue{5}b_5b_1 x_c^3 - \dgreen{5}b_5b_1^2)x_c + (1-b_1)x_c $ \\
      6 & $-b_6x_c^6 + \dblue{6}b_6b_1 x_c^4 - \dgreen{9}b_6b_1^2x_c^2 + \dred{2}b_6b_1^3+ (1-b_1)x_c $ \\
      7 & $-b_7x_c^7 + \dblue{7}b_7b_1 x_c^5 - \dgreen{14}b_7b_1^2x_c^3 + \dred{7}b_7b_1^3x_c + (1-b_1)x_c $ \\
      8 & $-b_8x_c^8 + \dblue{8}b_8b_1 x_c^6 - \dgreen{20}b_8b_1^2 x_c^4 + \dred{16}b_8b_1^3 x_c^2 - 2b_8b_1^4+ (1-b_1)x_c $ \\
      9 & $-b_9x_c^9 + \dblue{9}b_9b_1 x_c^7 - \dgreen{27}b_9b_1^2 x_c^5 + \dred{30}b_9b_1^3 x_c^3 + 9b_9b_1^4 x_c+ (1-b_1)x_c $ \\
      10 & $-b_{10}x_c^{10} + \dblue{10}b_{10}b_1 x_c^8 - \dgreen{35}b_{10}b_1^2 x_c^6 + \dred{50}b_{10}b_1^3 x_c^4 - 25 b_{10} b_1^4 x_c^2 + 2b_{10}b_1^5+ (1-b_1)x_c $ \\
      \hline
      & \strut $-b_nx_c^n + \dblue{n}b_nb_1 x_c^{n-2} - \dgreen{\frac{n(n-3)}{2}} b_nb_1^2 x_c^{n-4} + \dred{\frac{n(n-4)(n-5)}{6}} b_nb_1^3 x_c^{n-6} - \ldots + (1-b_1)x_c$ \\
    \end{tabular}
  \end{center} \vspace{-4mm}
  \vspace{2pt}
  \caption{Examples of the analytic relationship $Bt_c=Q(x_c)$ for the horizontal
  collapse point position $x_c$ that were inferred numerically using the
  high-precision calculations, for the case of an object described by two real
  non-zero Laurent coefficients $b_1$ and $b_n$. The integer coefficients
  colored in blue, green, and red follow patterns. The final line of the table
  shows an inferred general formula.\label{tab:anal1}}
\end{table*}

\begin{table*}
  \begin{center}
    \normalsize
    \begin{tabular}{l|l}
      $n$ & $Q(x_c)$ \\
      \hline
      3 & $-b_3x_c^3 + \dblue{3}b_3b_2 + (x_c - b_2 x_c^2)$ \\
      4 & $-b_4x_c^4 + \dblue{4}b_4b_2 x_c + (x_c - b_2 x_c^2)$  \\
      5 & $-b_5x_c^5 + \dblue{5}b_5b_2 x_c^2 + (x_c - b_2 x_c^2)$\\
      6 & $-b_6x_c^6 + \dblue{6}b_6b_2 x_c^3 - \dgreen{3}b_6b_2^2 + (x_c - b_2 x_c^2)$ \\
      7 & $-b_7x_c^7 + \dblue{7}b_7b_2 x_c^4 - \dgreen{7}b_7b_2^2 x_c + (x_c - b_2 x_c^2)$ \\
      8 & $-b_8x_c^8 + \dblue{8}b_8b_2 x_c^5 - \dgreen{12}b_8b_2^2 x_c^2 + (x_c - b_2 x_c^2)$ \\
      9 & $-b_9x_c^9 + \dblue{9}b_9b_2 x_c^6 - \dgreen{18}b_9b_2^2 x_c^3 + \dred{3}b_9b_1^3 + (x_c - b_2 x_c^2)$ \\
      10 & $-b_{10}x_c^{10} + \dblue{10}b_{10}b_2 x_c^7 - \dgreen{25}b_{10}b_2^2 x_c^4 + \dred{10}b_{10}b_2^3 x_c + (x_c - b_2 x_c^2)$ \\
      11 & $-b_{11}x_c^{11} + \dblue{11}b_{11}b_2 x_c^8 - \dgreen{33}b_{11}b_2^2 x_c^5 + \dred{22}b_{11}b_2^3 x_c^2 + (x_c - b_2 x_c^2)$ \\
      12 & $-b_{12}x_c^{12} + \dblue{12}b_{12}b_2 x_c^9 - \dgreen{42}b_{12}b_2^2 x_c^6 + \dred{40}b_{12}b_2^3 x_c^3 - 3 b_{12}b_2^4+ (x_c - b_2 x_c^2)$ \\
      \hline
      & \strut $-b_nx_c^n + \dblue{n}b_nb_2 x_c^{n-3} - \dgreen{\frac{n(n-5)}{2}} b_nb_2^2 x_c^{n-6} + \dred{\frac{n(n-7)(n-8)}{6}} b_nb_2^3 x_c^{n-9} - \ldots + (x_c - b_2 x_c^2)$ \\
    \end{tabular}
  \end{center} \vspace{-4mm}
  \vspace{2pt}
  \caption{Examples of the analytic relationship $Bt_c=Q(x_c)$ for the horizontal
  collapse point position $x_c$ that were inferred numerically using the
  high-precision calculations, for the case of an object described by two real
  non-zero Laurent coefficients $b_2$ and $b_n$. The integer coefficients
  colored in blue, green, and red follow patterns. The final line of the table
  shows an inferred general formula.\label{tab:anal2}}
\end{table*}

In Tables~\ref{tab:anal1} and \ref{tab:anal2}, we observe that each pair of
non-zero Laurent coefficients leads to a combination of additional terms
appearing in the collapse point polynomial. Building on these results, we
inferred and numerically tested the formula
\begin{align}
  Bt_c&= - b_4 x_c^4 - x_c^3 + x_c^2  (4 b_4 b_1 - b_2) \nonumber \\
  &\pheq + x_c (1 - b_1 + 3 b_3 b_1 + 4 b_4 b_2) \nonumber \\
  &\pheq + 2 b_2 b_1 + 3 b_3 b_2 + 4 b_4 b_3 - 2 b_4 b_1^2
  \label{eq:onumform}
\end{align}
for the case of all four coefficients $b_1$, $b_2$, $b_3$, and $b_4$ being
non-zero. In Eq.~\eqref{eq:onumform} all terms involve powers of two different
$b_n$, but for higher non-zero Laurent coefficients, terms with three or
more different $b_n$ arise. If $b_1$, $b_2$, and $b_5$ are non-zero, then we
find that
\begin{align}
  Bt_c &= - b_5 x_c^5 + 5 b_5 b_1 x_c^3 + (5b_5b_2 - b_2) x_c^2 \nonumber \\
  &\pheq + (1- b_1-5b_5b_1^2) + 2 b_2 b_1 - 4 b_5 b_2 b_1, \label{eq:five}
\end{align}
where the last term on the right hand side is a product of all three non-zero
Laurent coefficients.

The final generalization that we consider is when the Laurent coefficients are
complex. The fitting procedure described in Fig.~\ref{fig:fit_analytic} becomes
more complicated in this case, since both the horizontal position $x_c$ and the
vertical position $y_c$ of the collapse point will vary. For the case of the
Laurent coefficients $q_1$ and $q_4$ being non-zero and complex, we inferred
the formula
\begin{equation}
  \label{eq:numformula}
  Bt_c = \zb_c - \bq_1 z_c - \bq_4 z_c^4 + 4 \bq_4 q_1 z_c^2 - 2\bq_4 q_1^2,
\end{equation}
which is a generalization of the formula for $n=4$ in Table~\ref{tab:anal1}.
The generalization to complex coefficients introduces conjugates on some terms,
and the right hand side is not an analytic function of $z_c$ due to the first
term featuring $\zb_c$. If $q_4=0$ also, then Eq.~\eqref{eq:numformula}
simplifies to
\begin{equation}
  \frac{B}{2} = \frac{\zb_c - \bq_1 z_c}{1-q_1\bq_1},
\end{equation}
which is equivalent to the formula for an ellipse \smash{$z_c =
\frac{B}{2}(1+q_1)$} that was derived in previous work~\cite{bazant06a}.

\section{Derivation of the collapse point formulae}
\label{sec:deriv}
The previous section revealed a surprising and complicated connection between
the collapse point, initial shape of the object, and the flow strength. Using
these numerical results as a guide, we now analytically derive this connection.
While the formulae in Tables~\ref{tab:anal1} and \ref{tab:anal2} are
complicated, it is reasonable to imagine that the specific coefficients could
occur as the residue from a contour integral, perhaps involving the product
of several Laurent series, and thus our first step is to consider a general
integral quantity and determine its behavior during the dissolution process.

\subsection{Time-evolution of an integral quantity}
Consider the expression
\begin{equation}
  \label{eq:integral}
  I(t) = \oint_{S(t)} F(z) d\zb
\end{equation}
where $z=g(w)$, $S(t)$ is the shape of the object, and $F$ is an arbitrary
analytic function. This can be written as
\begin{equation}
  I(t) = \oint_C F(g(w)) \ob{g'(w)} d \wb,
\end{equation}
where $C$ is the unit circle, and $w=e^{i\theta}$. Since $\bar{w}w=1$ on the
unit circle, this can be converted into the integral of an analytic function,
\begin{equation}
  I(t) =-\oint_C F(g(w)) \bar{g}'\oow \frac{dw}{w^2},
\end{equation}
which can be written as
\begin{equation}
  I(t)=\oint_C F(g(w)) \frac{d}{dw} \left( \bar{g} \oow \right) dw
\end{equation}
and hence integration by parts can be used to obtain
\begin{equation}
  \label{eq:greg}
  I(t) = - \oint_C F'(g(w)) g'(w) \bar{g}\oow dw.
\end{equation}
This is the first of two expressions that will be used later. To obtain a
second integral expression, consider taking the time derivative, which gives
\begin{align}
  \frac{dI}{dt} &= -\oint_C \frac{d}{dw} \Big( F'(g(w))\Big) \dot{g}(w) \bar{g}\oow dw \nonumber \\
  &\pheq -\oint_C F'(g(w)) \left(\dot{g}'(w) \bar{g}\oow + g'(w) \dot{\bar{g}}\oow \right) dw.
  \label{eq:tder}
\end{align}
Applying integration by parts to the first integral, transfers the derivative
onto the $\dot{g}(w) \bar{g}(1/w)$ terms. Note that
\begin{equation}
\frac{d}{dw} \left(\dot{g}(w) \bar{g}\oow \right)= \dot{g}'(w) \bar{g}\oow - \frac{\dot{g}(w)}{w^2} \bar{g}'\oow
\end{equation}
and since the first term of this expression will cancel with one of the terms in
second integral of Eq.~\eqref{eq:tder}, it follows that
\begin{equation}
  \frac{dI}{dt} = - \oint_C F'(g(w)) \left( \frac{\dot{g}(w)}{w^2}\bar{g}'\oow + g'(w) \dot{\bar{g}}\oow \right) dw.
\end{equation}
By substituting $w=e^{i\theta}$ and making use of Eq.~\eqref{eq:time_ev},
\begin{align}
  \frac{dI}{dt} &= -\inttp F'(g(\ep)) i \Big(\dot{g}(\ep) \ob{g'(\ep)} \en \nonumber \\
  &\pheq
  + g'(\ep) \ob{\dot{g}(\ep)} \ep\Big) \,d\theta \nonumber \\
    &= -2i \inttp F'(g(\ep)) \Real \left(\ob{g'(\ep)\ep} \dot{g}(\ep)\right) \, d\theta \nonumber \\
    &= 2i \inttp F'(g(\ep)) (1-Ba\cos \theta ) \, d\theta.
\end{align}
This can be written as a contour integral as
\begin{align}
  \frac{dI}{dt} &= i \inttp F'(g(\ep)) (2 - Ba\ep - Ba \en) d\theta \nonumber \\
  &= \oint_C \frac{F'(g(w))(2w - Ba -Ba w^2) dw}{w^2},
  \label{eq:gtimeev}
\end{align}
yielding the second integral expression that will be used later. As a check, it
is useful to consider when $F'(z)=1$, in which case the integral matches the
one in Eq.~\eqref{eq:carea} for the area of the object. Then
\begin{equation}
\frac{dI}{dt} = \oint_C \frac{(2w - Ba -Baw^2) dw}{w^2} = 2\pi i (2) = 4\pi i
\end{equation}
and
\begin{align}
I(t) &= - \oint_C \left( a - \sum_{n=0}^N q_n n w^{-(n+1)}\right )\left( \frac{a}{w} + \sum_{n=0}^N \bq_n w^{n} \right) dw \nonumber \\
&= -2\pi i \left( a^2 - \sum_{n=0}^N n|q_n|^2 \right) = -2\pi i A(t)
\end{align}
This gives $\dot{A}(t)= -2$, which agrees with Eq.~\eqref{eq:ctime}.

\subsection{A specific integral quantity}
A interesting candidate for the function $F'$ is
\begin{equation}
  \label{eq:gform}
  F'(z) = \frac{1}{z-z_c}
\end{equation}
where $z_c$ is the collapse point. This function is particularly special, since
as the object is dissolving, the integrals given in Eqs.~\eqref{eq:greg} and
\eqref{eq:gtimeev} will always be finite, as the integration contour will never
pass over the singularity. Even though the function $F(z)=\log(z-z_c)$ is
multivalued, the two integral expressions that will be used in the following
derivation, Eqs.~\eqref{eq:greg} and \eqref{eq:gtimeev}, are related to each other
through a derivation only involving $F'$, and thus it is not necessary to
consider branch cuts that would be needed to integrate $F$.
Equation~\eqref{eq:gtimeev} will give
\begin{align}
  \frac{dI}{dt} &= \oint_C \frac{(2w- Ba - Baw^2) dw}{w^2 (aw - z_c + \sum_{n=0}^N q_n w^{-n} ) } \nonumber \\
  &= \oint_C \frac{(2w - Ba - Baw^2) dw}{aw^3\left(1 - \frac{1}{aw} \left(z_c-\sum_{n=0}^N q_n w^{-n} \right)\right) }.
\end{align}
Since $g(w,t)$ is a conformal map that takes the region $|w|\ge 1$ to the
region outside the object, there can be no solutions to $g(w,t)=z_c$ for
$|w|\ge 1$ and thus the above integrand will have no poles for $|w| \ge 1$.
Hence the integration contour can be deformed outwards and evaluated in terms
of the residue at infinity, which is given by the coefficient of the $w^{-1}$
term, namely $-Ba/a = -B$. Hence
\begin{equation}
  \frac{dI}{dt}= -2\pi i B
\end{equation}
and therefore $I(t)=D- 2\pi i Bt$ for some constant $D$. To determine the
constant, we consider the limit as $t\to t_c$, where $a\to 0$, $q_0 \to z_c$,
and $q_n\to 0$ for all $n>0$. Then Eq.~\eqref{eq:greg} shows that
\begin{equation}
  \label{eq:z_collapse}
  I(t) \to -\oint_C \frac{a\left( \frac{a}{w} + \bq_0 \right) dw}{(aw -z_c + q_0 ) }
  =-\oint_C \frac{a \bq_0}{aw} dw = -2\pi i \bar{z}_c
\end{equation}
and therefore $I(t)=2\pi i (B(t_c-t)-\bar{z}_c)$. Returning to Eq.~\eqref{eq:greg},
and examining the integral at the initial time,
\begin{equation}
  I(0) = - \oint_C \frac{\left( a - \sum_{n=0}^N q_n n w^{-(n+1)}\right )\left( \frac{a}{w} + \sum_{n=0}^N \bq_n w^{n} \right) dw}{aw -z_c + \sum_{n=0}^N q_n w^{-n} },
\end{equation}
which for large $w$ can be expanded as
\begin{widetext}
\begin{align}
  I(0) &= -\oint_C \frac{\left( a - \sum_{n=0}^N q_n n w^{-(n+1)}\right )\left( \frac{a}{w} + \sum_{n=0}^N \bq_n w^{n} \right) dw}{(aw- z_c + \sum_{n=0}^N q_n w^{-n} ) } \nonumber \\
  &= -\oint_C \frac{\left( a - \sum_{n=0}^N q_n n w^{-(n+1)}\right )\left( \frac{a}{w} + \sum_{n=0}^N \bq_n w^{n} \right) dw}{aw \left(1 - \frac{1}{aw} \left(z_c- \sum_{n=0}^N q_nw^{-n}\right) \right) } \nonumber \\
  &= -\oint_C \frac{1}{aw} \left( a - \sum_{n=0}^N \frac{nq_n}{w^{n+1}}\right) \left( \frac{a}{w} + \sum_{n=0}^N \bq_n w^{n} \right) \left( \sum_{k=0}^\infty \frac{\left(z_c- \sum_{n=0}^N q_nw^{-n}\right)^k}{a^k w^k}  \right) dw.
\end{align}
This integral will give the desired relationship between $z_c$ and $B$. By
expanding out the three power series, and looking at terms of the form $w^{-1}$
that will give a residue at infinity, the integral will simplify to a
polynomial in $z_c$. For example, consider the case of only $q_1$ and $q_4$
being non-zero. In that case, for $w$ large, and neglecting terms smaller than
$w^{-1}$,
\begin{align}
  I(0) &= -\oint_C \left( \frac{1}{w} - \frac{q_1}{aw^3} - \frac{4 q_4}{aw^6}\right) \left(\frac{a}{w} + \bq_1w + \bq_4w^4\right) \left( \sum_{k=0}^\infty \frac{1}{a^k}\left(\frac{z_c}{w} - \frac{q_1}{w^2} - \frac{q_4}{w^5} \right)^k\right) dw \nonumber \\
  &= -\oint_C \left( \bq_4 w^3 + \bq_1 - \frac{\bq_4 q_1w }{a} + \ldots \right) \left( 1 +  \frac{1}{a}\left(\frac{z_c}{w} - \frac{q_1}{w^2} - \frac{q_4}{w^5} \right) + \frac{1}{a^2}\left(\frac{z_c}{w} - \frac{q_1}{w^2} - \frac{q_4}{w^5} \right)^2 + \ldots \right) dw \nonumber \\
  &= -\oint_C \left( \frac{1}{w} \left( \frac{\bq_4 z_c^4}{a^4} - \frac{3\bq_4 q_1 z_c^2}{a^3} + \frac{\bq_4 q_1^2}{a^2} + \frac{\bq_1z_c}{a} - \frac{\bq_4 q_1 z_c^2}{a^3} + \frac{\bq_4 q_1^2}{a^2} \right) + \ldots \right) dw \nonumber \\
  &= -2\pi i \left( \frac{\bq_4 z_c^4}{a^4} - \frac{4\bq_4 q_1 z_c^2}{a^3} + \frac{2\bq_4 q_1^2}{a^2} + \frac{\bq_1z_c}{a}\right).
\end{align}
\end{widetext}
By using $I(0) = 2\pi i (B t_c-\bar{z}_c)$ it follows that
\begin{equation}
  a^4 Bt_c = a^4\bar{z}_c-\bq_4z_c^4 + 4a \bq_4 q_1 z_c^2- 2 a^2\bq_4q_1^2 - a^3\bq_1 z_c.
\end{equation}
If $a=1$ then
\begin{equation}
  \label{eq:case14}
  Bt_c = \zb_c-\bq_4z_c^4 + 4\bq_4 q_1 z_c^2 - 2 \bq_4q_1^2 - \bq_1 z_c,
\end{equation}
which agrees with Eq.~\eqref{eq:numformula} that was found numerically.

\begin{figure*}
  \begin{center}
    \normalsize
    \include{ex1_sp_evo}
    \setlength{\unitlength}{0.0150bp}
    \vspace{-0.1cm}
    \begin{picture}(21100,2800)(0,1000)
      \put(1600,2000){\includegraphics[width=271pt,height=15pt]{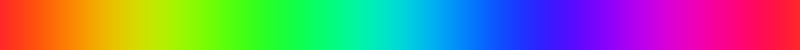}}
      \put(1600,2000){\line(1,0){18000}}
      \put(1600,1800){\line(0,1){1200}}
      \put(1600,3000){\line(1,0){18000}}
      \put(19600,1800){\line(0,1){1200}}
      \put(6100,1800){\line(0,1){200}}
      \put(10600,1800){\line(0,1){200}}
      \put(15100,1800){\line(0,1){200}}
      \put(1600,1400){\makebox(0,0)[c]{0}}
      \put(6100,1400){\makebox(0,0)[c]{$\pi/2$}}
      \put(10600,1400){\makebox(0,0)[c]{$\pi$}}
      \put(15100,1400){\makebox(0,0)[c]{$3\pi/2$}}
      \put(19600,1400){\makebox(0,0)[c]{$2\pi$}}
    \end{picture}
  \end{center} \vspace{-4mm}
  \caption{(a) The thick black line shows an first example object, where $a=1$
  and the only non-zero Laurent coefficients are
  \smash{$q_1=\frac{1}{10}+\frac{3}{20}i$} and
  \smash{$q_4=\frac{1}{10}+\frac{1}{20}i$}. The colors show the argument of the
  function $P(z)$ for \smash{$B=\frac{7}{20}$}, whose roots represent
  candidates for the collapse point of the object as it dissolves. The dashed
  lines are contours of $|P(z)|$ at
the
  values of \smash{$\frac{n^2-n+1}{2}$} for
  $n\in \N$. (b) A zoomed-in region showing forward and backward time-evolution
  of the object boundary at intervals of \smash{$\frac{1}{20}t_c$}. The unique
  negative-sense root of $P$ is shown by a circle, and one of the
  positive-sense roots is shown by a triangle. The four other positive-sense
  roots are outside the region that is plotted.\label{fig:ex1_sp_evo}}
\end{figure*}
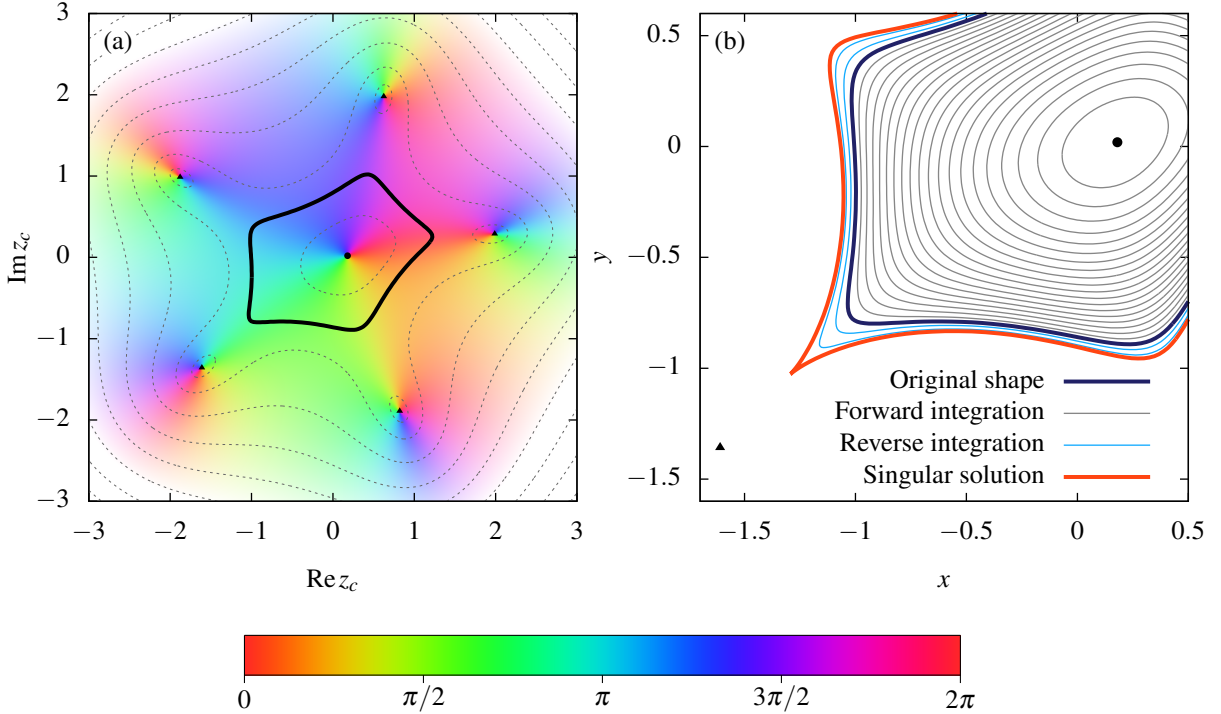

\section{Three examples of the collapse point equation}
\label{sec:examples}
For a general case, the collapse point $z_c$ satisfies the equation
\begin{equation}
  \label{eq:solp}
  0 = P(z_c) = \zb_c - Bt_c + \frac{1}{2\pi i} \oint_{S(0)} \log (z-z_c) d\zb,
\end{equation}
defined at points within the object, where the integral in this equation is
evaluated as a polynomial in $z_c$ following the series expansion procedure
described in the previous section. The polynomial can then be analytically
extended to give an expression for $P(z_c)$ at points outside the object also.
However, at points outside the object, the analytic extension will not match
the value of the integral, since the enclosed residues will be different.
Equation~\eqref{eq:solp} is complicated: it is not analytic due to the presence
of $\zb_c$, and in general it will contain higher powers of $z_c$, so it is
likely to have multiple solutions. To use this equation as a predictive tool,
it is useful to understand the typical structure of $P$ and know how to select
the correct root. We now consider three examples that explore the structure of
$P$ in relation to the object shape.

\subsection{First example: an irregular pentagonal shape}
\label{sub:ex1}
Consider an example based on Eq.~\eqref{eq:case14}, where the function $P$ can be
written as
\begin{equation}
P(z_c) = \zb_c - B t_c - \bq_4z_c^4 + 4\bq_4 q_1 z_c^2 - 2 \bq_4q_1^2 - \bq_1 z_c,
\end{equation}
where $t_c=1-|q_1|^2-4|q_4|^2$. Fig.~\ref{fig:ex1_sp_evo} shows a plot of the
modulus and argument of this function for the case of $a=1$,
\smash{$q_1=\frac{1}{10}+\frac{3}{20}i$},
\smash{$q_4=\frac{1}{10}+\frac{1}{20}i$}, and \smash{$B=\frac{7}{20}$}. The
shape of the object is also shown. There are five roots that lie outside the
object. There is one root inside the object, which must be the collapse point.
Furthermore, the argument in the neighborhood of the interior root rotates in
the negative (anti-analytic) sense, whereas the argument near each exterior
root rotates in the positive (analytic) sense. By considering the Taylor series
of $P$ at a given root, one can mathematically determine whether a root is
positive-sense or negative-sense by whether $|P_z|^2-|P_{\zb}|^2$ is positive
or negative, respectively. From Eq.~\eqref{eq:z_collapse}, the collapse point
must be given by a negative-sense root, and hence for this example there is an
unambiguous choice, of the single negative-sense root within the object.

It is interesting to consider whether the other roots have physical
significance. Figure~\ref{fig:ex1_sp_evo}(b) shows a zoomed-in region of the
dissolution process for this example, confirming that the interior
negative-sense root visible in Fig.~\ref{fig:ex1_sp_evo}(a) is indeed the
collapse point. The figure also shows a nearby positive-sense root. If the
system is time-integrated backward, then the boundary of the object sharpens
toward the root. This leads to a cusp singularity in a finite time
$t=-0.06133$, which appears similar to cusp development in related
systems~\cite{shraiman84,howison86,tanveer00}. The cusp formation occurs when
a branch point of $g$ reaches the unit circle. As the cusp is approached, the
matrix $M(\vec{s})$ becomes singular, and the \textsc{DOP853} integrator
terminates because the timestep required to keep the local error below the
tolerance is smaller than what can be resolved with double precision. While the
positive-sense root appears connected to the development of the cusp, it is not
located exactly at the cusp, and thus it is not clear what, if any, its precise
physical significance is.

\begin{figure*}
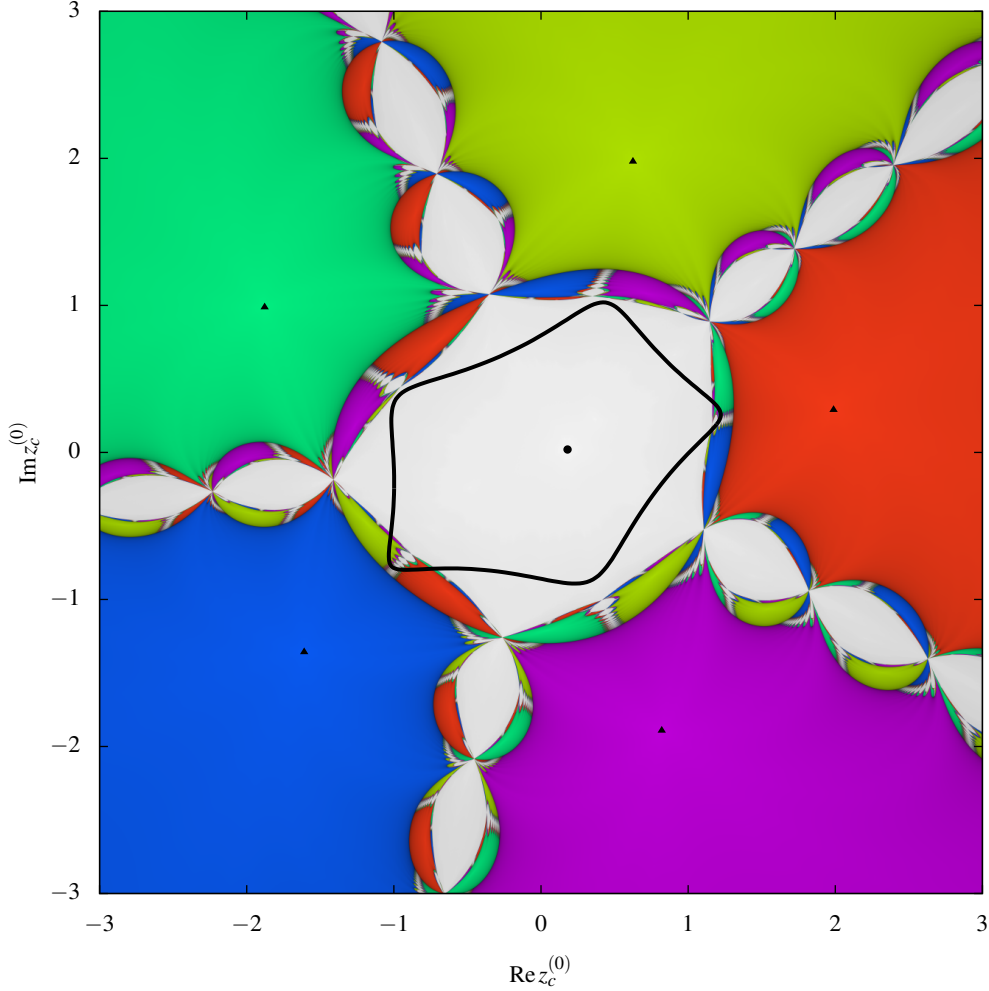

  \begin{center}
    \include{ex1_newton}
  \end{center} \vspace{-4mm}
  \caption{Plot showing which root of $P$ a Newton--Raphson iteration will
  converge to when starting at \smash{$z_c^{(0)}$}, for the example
  configuration given in Fig.~\ref{fig:ex1_sp_evo}. The five positive-sense roots
  of $P$ are shown by small black triangles, and the unique negative-sense root
  is shown by a small black circle. Each point is colored according to the
  argument of the root that it converges to, with the central root being shown
  in white. Darker shades show regions that require more iterations to
  converge.\label{fig:ex1_newton}}
\end{figure*}

A practical way to determine the root positions is to make use of a
Newton--Raphson iteration, generalized to take into account that $P$ also
depends on the conjugate of $z_c$. An appropriate Newton--Raphson iteration can
be constructed by viewing $P$ as a function of two variables $z_c$ and $\zb_c$,
and considering the two-function system of $P$ and $\Pb$. For a guess of the
form \smash{$z_c^{(n)}$}, the vector generalization of the Newton--Raphson
method to give an improved guess \smash{$z_c^{(n+1)}$} is then
\begin{equation}
\left(
\begin{array}{cc}
  P_z & P_{\zb} \\
  \Pb_z & \Pb_{\zb}
\end{array}
\right)
\left(
\begin{array}{c}
  z_c^{(n+1)} - z_c^{(n)} \\
  \zb_c^{(n+1)} - \zb_c^{(n)}
\end{array}
\right)
=
-
\left(
\begin{array}{c}
  P \\ \Pb
\end{array}
\right),
\end{equation}
which leads to the two equations
\begin{align}
  P_z (z_c^{(n+1)} - z_c^{(n)}) + P_{\zb} (\zb_c^{(n+1)} - z_c^{(n)}) &= -P, \label{eq:nr1} \\
  \Pb_z (z_c^{(n+1)} - z_c^{(n)}) + \Pb_{\zb} (\zb_c^{(n+1)} - z_c^{(n)}) &= -\Pb. \label{eq:nr2}
\end{align}
Substituting Eq.~\eqref{eq:nr2} into Eq.~\eqref{eq:nr1} to eliminate
\smash{$(\zb_c^{(n+1)}-\zb_c^{(n)})$} gives the iterative equation
\begin{equation}
  \label{eq:newtraph}
  z_c^{(n+1)}=z_c^{(n)} + \frac{\Pb P_{\zb} - P \overline{P_z}}{|P_z|^2 - |P_{\zb}|^2}.
\end{equation}
As expected, if $P_{\zb}=0$, then this equation reduces to the standard complex
Newton--Raphson iteration.

\begin{figure*}
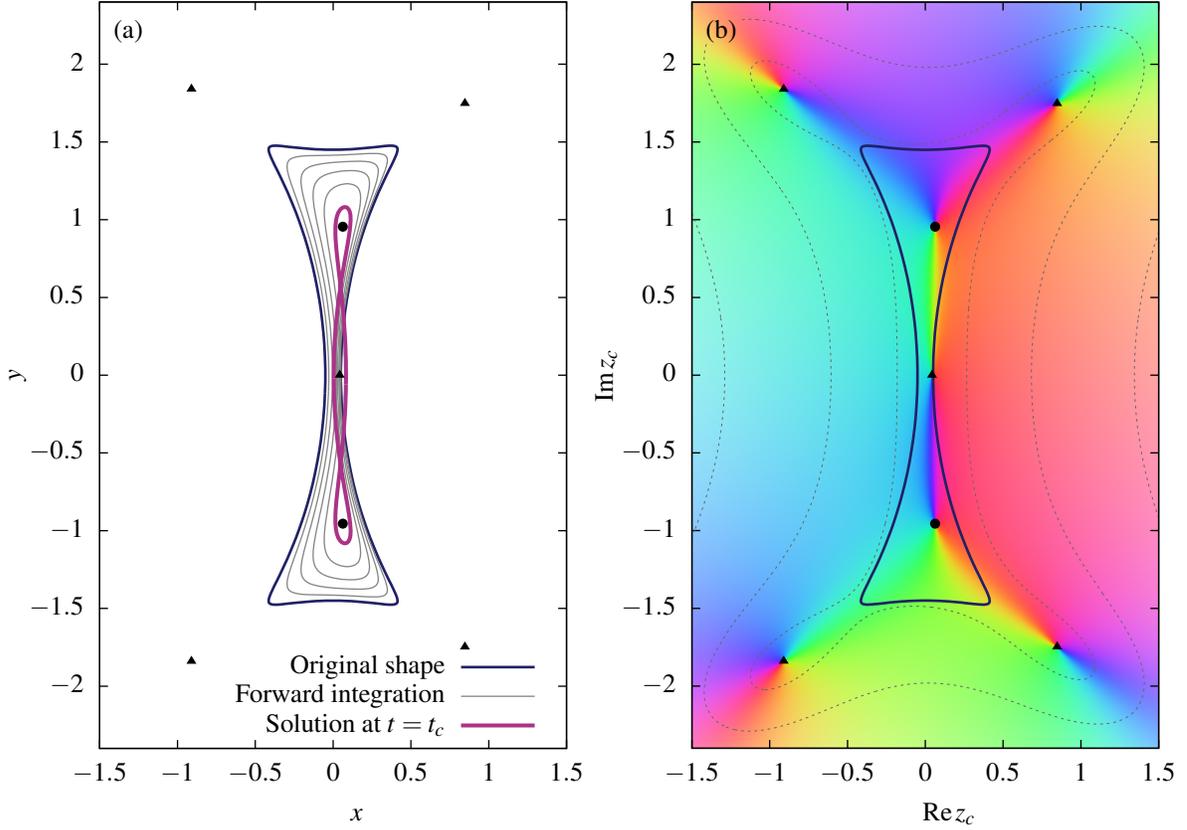

  \begin{center}
    {\normalsize
    \include{ex2_evo_sp}}
  \end{center} \vspace{-4mm}
  \vspace{-10pt}
  \caption{(a) Time-evolution of a dumbbell-shaped object described by $a=1$,
  \smash{$q_1=-\frac{7}{10}$}, \smash{$q_3=-\frac{1}{4}$},
  \smash{$B=\frac{3}{5}$}, and all other Laurent series coefficients are zero.
  The gray curves are plotted at intervals of \smash{$\frac{1}{5}t_c$}.
  Positive-sense roots of the function $P$ are shown by triangles, and
  negative-sense roots are shown by circles. (b) The structure of the
  corresponding function $P$, with the dashed lines corresponding to contours
  of $|P(z)|$ at \smash{$\frac{n^2-n+1}{4}$} for $n\in \N$, and the colors
  corresponding to the argument using the key given in
  Fig.~\ref{fig:ex1_sp_evo}.\label{fig:ex2_evo_sp}}
\end{figure*}

Figure~\ref{fig:ex1_newton} shows a plot of which root the Newton--Raphson
iteration will converge to as a function of the starting guess
\smash{$z_c^{(0)}$}. As is typical for Newton--Raphson iterations of complex
functions, the plot has a fractal structure, with large basins of attraction
surrounding each root. However, the plot has some distinctly different features
to usual Newton fractals~\cite{mandelbrot_book,peitgen_book} arising from the
vector generalization of the iteration to non-analytic functions. In
particular, the denominator $|P_z|^2 - |P_{\zb}|^2$ featuring in
Eq.~\eqref{eq:newtraph} is zero on a one-dimensional loop of points surrounding
the central root. Any starting guess that approaches this loop will therefore
undergo a very large initial step. In Fig.~\ref{fig:ex1_newton}, this loop
forms the dividing line between the five outer colored basins and the central
region. Due to the self-similarity of the fractal, the structure surrounding
this loop is replicated in other parts of the plot. This is in noticeable
contrast to the regular Newton fractal for an analytic function $f(z)$, where
the iteration becomes singular only at a zero-dimensional set of points where
$f'(z)=0$.

On Figure~\ref{fig:ex1_newton}, the object boundary is shown by the dashed
black line, and it is almost entirely contained within the central white
region, meaning that a starting guess within the object is likely to converge
to the collapse point; if the guess is chosen near the center of the object,
such as at $q_0$, the iteration converges very rapidly and reliably. However
the plot also indicates that for several small regions inside the object ({\it
e.g.} near the bottom left corner) the Newton--Raphson method may converge to
one of the exterior roots.

\subsection{Second example: a dumbbell-shaped object dividing in two}
\label{sub:ex2}
Figure~\ref{fig:ex2_evo_sp}(a) shows the dissolution process for the case of a
long dumbbell-shaped object, where $a=1$, \smash{$q_1=-\frac{7}{10}$},
\smash{$q_3=-\frac{1}{4}$}, \smash{$B=\frac{3}{5}$}, and all other Laurent
series coefficients are zero. In this case, the thin vertical sliver dissolves
away leaving two separated fragments. While the system can be time-integrated
past this point with $M(\vec{s})$ remaining non-singular, the contour begins to
overlap with itself, thus losing physical validity. Mathematically, this
scenario corresponds to when the function $g$ from the unit disk to the
physical domain becomes multivalued. This is a global property of the function
and is therefore a different type of finite-time singularity from the cusp
considered in the previous example.

The function $P$ for this example is
\[
P(z_c) = \zb_c - B t_c - \bq_3z_c^3 + 3\bq_3 q_1 z_c - \bq_1 z_c,
\]
where $t_c=1-|q_1|^2-3|q_3|$. The structure of $P(z_c)$ and its roots are
plotted in Fig.~\ref{fig:ex2_evo_sp}(b). The function $P$ has two
negative-sense roots in either end of the dumbbell, four exterior
positive-sense roots, and one positive-sense root on the vertical sliver. This
example also highlights that the non-analyticity of $P$ significantly increases
its complexity. The last four terms of $P$ form an analytic cubic function in
$z_c$, which could have at most three distinct roots, but adding the
anti-analytic $\zb_c$ increases the number of roots to seven.

\begin{figure}
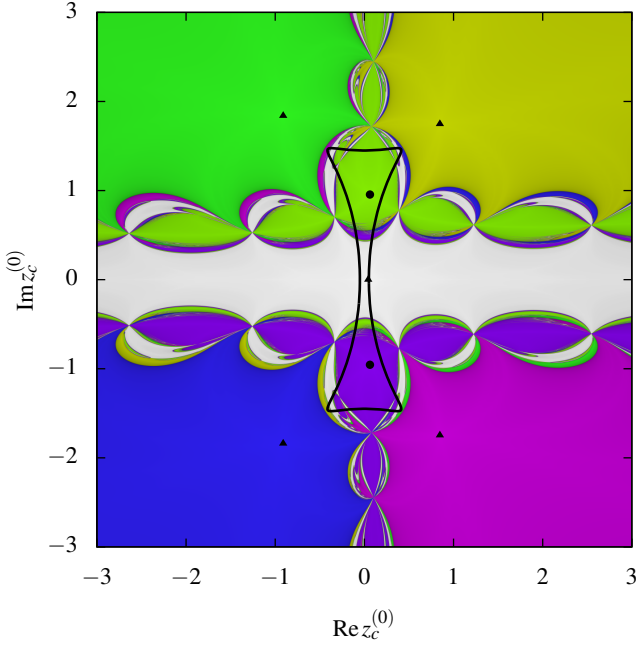
 
  \begin{center}
    \include{ex2_newton_prx}
  \end{center} \vspace{-4mm}
  \caption{Plot showing which root of $P$ a Newton--Raphson iteration will
  converge to when starting at \smash{$z_c^{(0)}$}, for the dumbbell-shaped
  example. The five positive-sense roots of $P$ are shown by black triangles,
  and the two negative-sense roots are shown by black circles. Each point is
  colored according to the argument of the root that it converges to, with the
  central root being shown in white. Darker shades show regions that require
  more iterations to converge.\label{fig:tfract}}
\end{figure} 

While more complicated than the previous example, the positions of the roots
appear to be physically reasonable, with one negative-sense root appearing in
each end of the dumbbell. The central positive-sense root appears to be
associated with the position where the vertical sliver dissolves. However,
close inspection reveals that its position is not perfectly aligned with the
point where the two sides of the object first come into contact. Instead, it
appears to mark the center of the inverted section of the contour at $t=t_c$.
Figure~\ref{fig:tfract} shows a plot of which root the Newton--Raphson
iteration converges to, depending on the starting guess. For starting points
\smash{$z_c^{(0)}$} within the object, most will converge to the two
negative-sense roots or the central positive-sense root. The denominator
\smash{$|P_z|^2 - |P_{\zb}|^2$} in Eq.~\eqref{eq:newtraph} vanishes on two
approximate ellipses surrounding each negative-sense root.

While it is not physically valid to simulate the dissolution of the object to
collapse, this example highlights that the structure of $P$ and the position of
its roots may be more complicated than in the previous example considered, and
thus any further mathematical analyses would have to take into account this
possibility.

\begin{figure*}
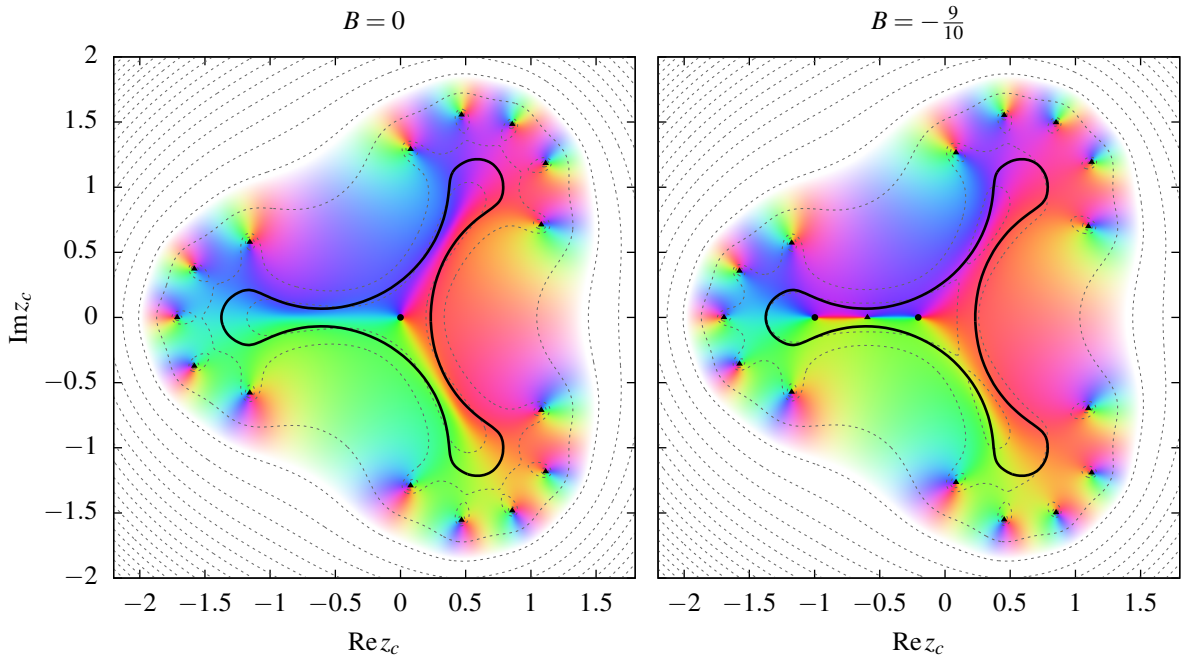

  \begin{center}
    {\normalsize
    \include{ex3_sptrans}}
  \end{center} \vspace{-4mm}
  \vspace{-3mm}
  \caption{Plot of the collapse point equation $P(z_c)$ for the three-pronged
  object for (a) $B=0$, and (b) \smash{$B=-\frac{9}{10}$}. The colors
  correspond to the argument of $P(z)$ using the same scale as
  Fig.~\ref{fig:ex1_sp_evo}. The positive-sense roots are shown with black
  triangles and the negative-sense roots are shown with black circles. The
  thick black line shows the boundary of the object. The thin dashed gray lines
  are contours of $|P(z)|$ at
  \smash{$\frac{1}{4}(n^4+1)$} for $n \in \N_0$.\label{fig:ex3_sptrans}}
\end{figure*}

\subsection{Third example: transitions in behavior as flow strength is altered}
\label{sub:ex3}
The final example is a three-pronged object given by the initial non-zero
Laurent coefficients $a=1$, \smash{$q_2=-\frac{49}{100}$},
\smash{$q_5=-\frac{17}{100}$}, \smash{$q_8=-\frac{3}{40}$},
\smash{$q_{11}=-\frac{27}{1000}$}, and \smash{$q_{14}=-\frac{3}{500}$}.
Unlike the previous two examples, the collapse point equation is difficult
to determine manually due to the large number of Laurent series terms
that must be considered. However, a computer code was written that found
it to be
\begin{align}
  P(z_c) &= \zb_c-Bt_c -\bq_{14} z_c^{14} + z_c^{11}(- \bq_{11} + 14 \bq_{14} q_2 ) \nonumber\\
  &\pheq+ z_c^8 ( -\bq_8 +11 \bq_{11} q_2 + 14 \bq_{14} q_5 - 63 \bq_{14} q_2^2) \nonumber \\
  &\pheq+ z_c^5 ( -\bq_5 + 8 \bq_8 q_2 + 11 \bq_{11} q_5 + 14 \bq_{14} q_8 \nonumber \\
  &\pheq- 33 \bq_{11} q_2^2 -84 \bq_{14} q_2q_5 + 98 \bq_{14} q_2^3) \nonumber \\
  &\pheq+ z_c^2 ( -\bq_2 + 5 \bq_5 q_2 + 8 \bq_8 q_5 + 11 \bq_{11} q_8 \nonumber \\
  &\pheq+ 14 \bq_{14} q_{11} - 12 \bq_8q_2^2 -32 \bq_{11} q_2 q_5 \nonumber \\
  &\pheq-42 \bq_{14} q_2 q_8 - 21 \bq_{14} q_5^2 + 22 \bq_{11} q_2^3 \nonumber \\
  &\pheq+ 84 \bq_{14} q_2^2 q_5 - 35 \bq_{14} q_2^4 ).
\end{align}
The left panel of Fig.~\ref{fig:ex3_sptrans} shows the structure of the
solution polynomial when $B=0$. Each of the three prongs is surrounded by five
positive-sense roots, and there is a single negative-sense root at the origin.
The magnitude of $P$ within the object is small, so that most of the object
lies within the region \smash{$|P(z_c)|<\frac{1}{4}$}, meaning that an
alteration of the flow strength could alter the function's roots. The right
panel shows the function $P$ when \smash{$B=-\frac{9}{10}$}, corresponding to a
strong flow from the right. In this case a new pair of positive-sense and
negative-sense roots appear on the real axis, resulting in a similar root
arrangement to Fig.~\ref{fig:ex2_evo_sp}(b).

\begin{figure}
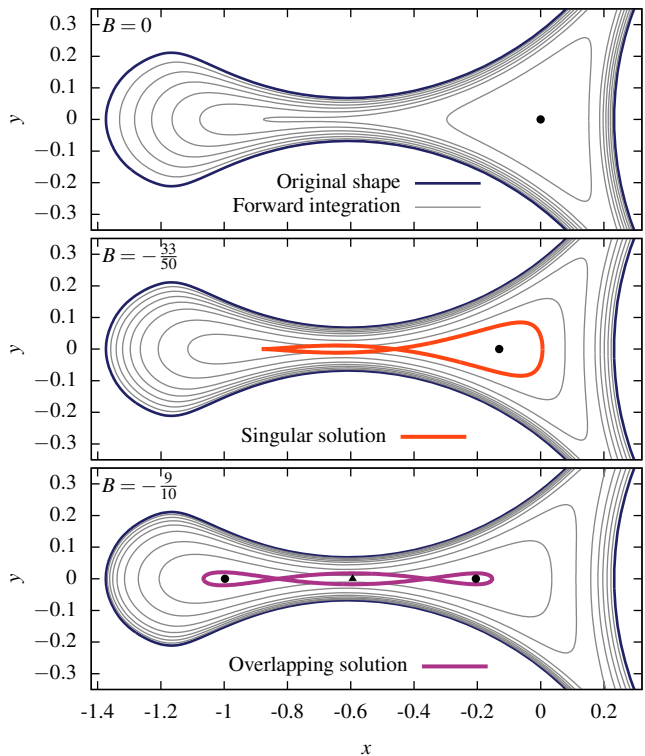
 
  \begin{center}
    {\footnotesize
    \include{ex3_prongs_prx}}
  \end{center} \vspace{-4mm}
  \caption{Zoomed-in plot of the dissolution process for the three-pronged
  object, for three different values of the flow strength $B$, showing
  snapshots of the object boundary at intervals of \smash{$\frac{t_c}{8}$}
  where $t_c=0.1608885$. The positive-sense roots are shown with black triangles
  and the negative-sense roots are shown with black circles. For $B=0$, the
  object collapses to the origin at $t=t_c$. For \smash{$B=-\frac{33}{50}$},
  the object boundary overlaps and then a singular solution with a cusp forms
  at $t=0.1550602$. For \smash{$B=-\frac{9}{10}$} the boundary forms an
  overlapping curve at $t=t_c$.\label{fig:ex3_prongs}}
\end{figure} 

Figure~\ref{fig:ex3_prongs} shows the dissolution process for three different
cases of $B$, for a zoomed-in region centered on one of the prongs. The top
panel shows the case when $B=0$, where the dissolution process proceeds
normally and the object collapses at the single negative-sense root at the
origin. Since the physical model given in Eqs.~\eqref{eq:confeq1},
\eqref{eq:confeq2}, \& \eqref{eq:ad_bc} tends to rapidly dissolve sharply curved
boundaries, the three prongs of the object dissolve rapidly enough that they
remain connected to each other.

For the case of \smash{$B=\frac{9}{10}$} shown in the bottom panel of
Fig.~\ref{fig:ex3_prongs} the situation is different. The incorporation of flow
into the evolution equation of Eq.~\eqref{eq:time_ev} causes the thin part of
the prong to dissolve more rapidly than its end, meaning that in this case, the
object becomes disconnected into two regions. The behavior is similar to the
previous example, where the object boundary overlaps with itself. At $t=t_c$,
the object boundary loops around the two negative-sense roots and the
positive-sense root in the same manner as Fig.~\ref{fig:ex2_evo_sp}(a). This
example highlights that only altering flow strength is sufficient to cause a
transition in the behavior of the dissolution process.

The transition in behavior is linked with the formation of the new roots in
Fig.~\ref{fig:ex3_sptrans} as $B$ is changed from $0$ to
\smash{$-\frac{9}{10}$}. However, the middle panel of Fig.~\ref{fig:ex3_prongs}
for an intermediate flow strength of \smash{$B=-\frac{33}{50}$} shows that this
transition is more complicated. In this case, there is only a single
negative-sense root in $P$. However, during time-evolution, the object boundary
first overlaps with itself, and then the left loop shrinks to zero size,
leading to a singular solution with an inverted cusp at time $t=0.1550602<t_c$.

We carried out a systematic sweep over the flow strengths over the range from
$B=0$ to $B=-1$: initially the object collapses to a single point, at $B
\approx -0.233$ an inverted cusp forms, and at $B \approx -0.794$ a second
negative-sense root forms, when the left loop is large enough to persist until
$t_c$. This result highlights that dissolution process can transition between
at least three distinct behaviors. Furthermore, the result for
\smash{$B=-\frac{33}{50}$} shows that even if $P$ only has a single
negative-sense root, the dissolution process may not be straightforward, and
may lead to an overlapping boundary or a singular solution.

\begin{figure}
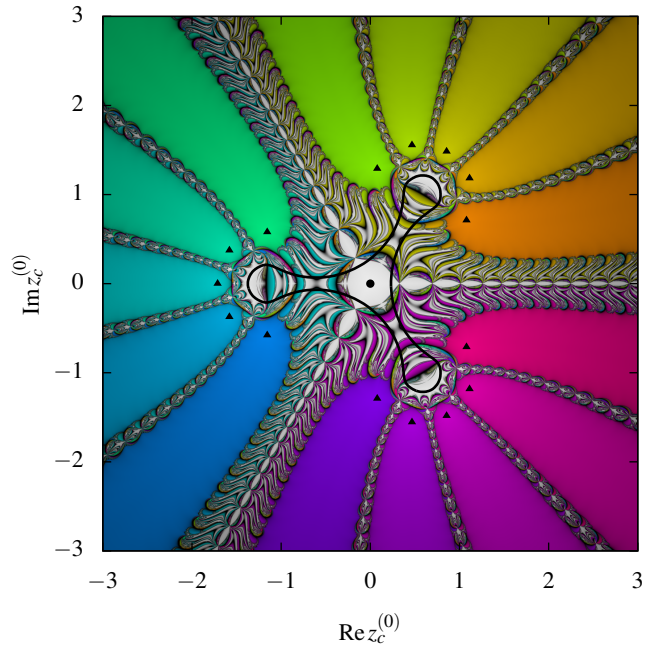
 
  \begin{center}
    \include{ex3_newton_prx}
  \end{center} \vspace{-4mm}
  \caption{Plot showing which root of the solution polynomial $P$ a
  Newton--Raphson iteration will converge to when starting at
  \smash{$z_c^{(0)}$}, for the three-pronged object when $B=0$. The fifteen
  positive-sense roots of $P$ are shown by black triangles, and the single
  negative-sense root at the origin is shown by a black circle. Each point is
  colored according to the argument of the root that it converges to, with the
  central root being shown in white. Darker shades show regions that require
  more iterations to converge.\label{fig:ex3_newton}}
\end{figure} 

Figure~\ref{fig:ex3_newton} shows which roots the generalized Newton--Raphson
iteration will converge to, for the case of $B=0$. The plot has an intricate
structure and there are many small, distinct regions that converge to the
central root. The denominator $|P_z|^2 - |P_{\zb}|^2$ in Eq.~\eqref{eq:newtraph}
vanishes on a small loop surrounding the central root, and starting guesses
near this loop are colored in darker shades, indicating that the root-finding
algorithm takes many iterations to converge. The plot highlights the difficulty
of finding the particular roots of interest in a general case.

\section{Conclusion}
\label{sec:conc}
In this paper, we studied a model of object dissolution within a
two-dimensional potential flow, and we created a numerical implementation of it
that allowed us to simulate the dissolution process for arbitrary objects
described in terms of a Laurent series. The simulations revealed an exact
relationship where the collapse point $z_c$ is the root of a non-analytic
function given in the terms of the Laurent coefficients and the flow strength.
This relationship was subsequently derived analytically, but it is unlikely
that it would have been discovered without the numerical results as a guide.
These simulations made use of a high-order numerical method, and while these
methods are often difficult or too computationally expensive to apply to real
engineering problems, this work demonstrates their power in mathematical
analysis: the numerical results for the collapse point are accurate enough to
infer the underlying exact relationship with reasonable confidence. There are
other examples where high-accuracy numerical methods have been used for
similar purposes, such as demonstrating the existence of special solutions to
equations~\cite{wilkening11,rycroft13} or to discovering universal
behavior~\cite{barenblatt14}.

The examples of Section~\ref{sec:examples} create some interesting theoretical
questions for future investigation. We expect that the first example of the
pentagonal shape (Subsec.~\ref{sub:ex1}) represents typical behavior for a
broad class of objects, where the dissolution process is well-defined, the
object collapses to a single point in finite time, and the collapse point
function $P(z_c)$ has a single negative-sense root. More specifically, we
expect this to be true for a large class of cases where the $q_j$ are small in
comparison to $a$, and hence the collapse point function in Eq.~\eqref{eq:solp}
will be dominated by the $\zb_c$ term and thus likely to have a single
anti-analytic root close to the origin. However, the second example shows that
not all cases may lead to this typical behavior, and object may dissolve into
multiple fragments, with $P(z_c)$ gaining additional negative-sense roots. The
third example adds a further complication, showing that only a minor alteration
of the flow strength $B$ can lead to cases where dissolution is not
well-defined, even though $P(z_c)$ still has a single negative-sense root.
The roots of $P(z_c)$ are connected to the formation of the finite-time
singularities, and give both an indication of the formation of local cusps, and
potential global topological changes. The cusp formation is similar to
continuous Laplacian growth~\cite{shraiman84,bensimon86} although a key
difference here is the breakage of symmetry due to the flow. This is
particularly well illustrated by the third example, where the addition of flow
breaks the three-fold symmetry and causes several transitions in behavior.

The collapse point results motivate two further questions: (A) what the
conditions on the initial modes for $P$ to have a single anti-analytic root,
and (B) what are the conditions on the initial modes for the dissolution
process to be well-defined and for the object to collapse to a single point?
If questions A and B can be answered, then a further direction would be to
identify a procedure capable of determining the collapse point with absolute
certainty. The generalized Newton--Raphson method that was introduced in
Subsec.~\ref{sub:ex1} is very efficient at identifying roots, but it is
difficult to determine {\it a priori} which root it will converge to, and plots
of the convergence as a function of the starting guess exhibit a fractal
structure as is typical for complex Newton--Raphson iterations. Furthermore,
the non-analyticity of $P(z_c)$ creates some difficulties whereby the total
number of roots exhibits fundamentally different behavior than for analytic
functions. An analytic cubic polynomial in $z_c$ has exactly three roots (when
counted with multiplicity) but the addition of a non-analytic $\zb_c$ as in the
second example (Subsec.~\ref{sub:ex2}) leads to seven roots, five of which are
positive-sense and two of which are negative-sense. By using a winding
argument, considering the curve $P(Re^{i\theta})$ as $R\to \infty$, we obtain
$n_+-n_-=N$ where $N$ is the maximum non-zero mode, and $n_\pm$ are the number
of positive-sense and negative-sense roots. We also consider the
Newton--Raphson fractals to be interesting in their own right, since they have
a fundamentally different structure than typical Newton--Raphson fractals due
to the one-dimensional set of points where the denominator in
Eq.~\eqref{eq:newtraph} vanishes. There is an interesting correspondence whereby
each object has an associated fractal.

A variety of generalizations to the dissolution model could also be explored.
The simple form of the right hand side of Eq.~\eqref{eq:time_ev} was based on
asymptotic considerations of the concentration profile in the low P\'eclet
number limit, but the numerical method could be extended to more complex growth
laws where higher powers of $\cos \theta$ and $\sin \theta$ are present. Since
the derivation of the collapse point function is not highly dependent on the
simple form of Eq.~\eqref{eq:time_ev}, it may be possible to generalize this to
more complex growth laws as well. Another extension is to the case of regular
polyhedral objects, which could be approximated using many terms in a Laurent
series.

The second and third examples of Subsecs.~\ref{sub:ex2} and \ref{sub:ex3} show
that in some cases an object may dissolve into several components. In the
current numerical method the dissolution process cannot be accurately simulated
beyond the point where multiple fragments form, but it may be possible to extend
the simulation to this case by using recent advances in conformal mapping for
multiply connected domains~\cite{delillo04,crowdy05,delillo06,crowdy07}. The
dissolution model is a particularly interesting example, since the physical
process involves a single domain smoothly transitioning into two. We aim to
investigate all of these interesting directions in future work.

Finally, we mention the discrete, stochastic analog of this problem. In the
absence of advection, the diffusion-limited erosion of a surface leads to
smooth, stable evolution that resembles the continuum limit of
diffusion-limited dissolution~\cite{paterson84,tang85,meakin86,krug91}, but to
our knowledge this model has never been analyzed (or even simulated) to the
point where the last particle is removed. In the final stages of collapse,
discreteness must again become important. Of course, the same applies to
advection--diffusion-limited dissolution, or any other conformally invariant
dissolution model~\cite{bazant03,bazant06a}. We thus leave the reader with an
open question: what is the probability that a given particle is the last to be
removed by advection--diffusion-limited erosion of a finite cluster in a fluid
flow? 

\section*{Acknowledgments}
C.~H.~Rycroft thanks Jue Chen (University of California, Berkeley) for useful
discussions. C.~H.~Rycroft was supported by the Director, Office of Science,
Computational and Technology Research, U.S. Department of Energy under contract
number DE-AC02-05CH11231.

\appendix
\section{Component form of the time-evolution equation}
\label{app:numerics}
The numerical method introduced in Section~\ref{sec:numerics} is based
upon equating the different sine and cosine components of Eq.~\eqref{eq:pgcomp},
which is
\begin{align*}
-1+Ba\cos\theta &= \Real\Bigg(\left[ a e^{-i\theta} - \sum_{n=0}^N n
	(b_n-ic_n) e^{in\theta}\right] \\
	&\pheq \left[\dot{a} e^{i\theta} + \sum_{n=0}^N
	(\dot{b}_n + i \dot{c}_n) e^{-in\theta}\right]\Bigg).
\end{align*}
Multiplying out these power series yields
\begin{align}
  -1 + Ba \cos \theta &= \Real\bigg( a\dot{a} - \dot{a} \sum_{n=0}^N n(b_n - i c_n)
	e^{i(n+1)\theta} \nonumber \\
	& \pheq
	+ a \sum_{m=0}^N (\dot{b}_m + i\dot{c}_m) e^{i(m+1)\theta} \nonumber \\
	&\pheq - \sum_{n=0}^N \sum_{m=0}^N n (b_n-ic_n)(\dot{b}_m+i\dot{c}_m)
	e^{i(n-m)\theta}\bigg).
\end{align}
Taking the real component of the bracketed term yields
\begin{align}
 -1 + Ba \cos \theta &= a\dot{a} - \sum_{n=0}^N \sum_{m=0}^N n\Big[(b_n\dot{b}_m + c_n \dot{c}_m)
	\cos (n-m)\theta
	\nonumber \\ & \pheq
	+ (c_n \dot{b}_m- b_n\dot{c}_m) \sin (n-m)\theta
	\Big] \nonumber \\
	& \pheq+ a \sum_{m=0}^N (\dot{b}_m \cos (m+1)\theta + \dot{c}_m
	\sin (m+1)\theta) \nonumber \\
	& \pheq -\dot{a} \sum_{n=0}^N (b_n \cos (n+1)\theta + c_n \sin
	(n+1)\theta ).
\end{align}
Collecting terms with factors of sine and cosine yields
\begin{align}
-1 + Ba \cos \theta &= a\dot{a} - \sum_{n=0}^N n (b_n \dot{b}_n + c_n \dot{c}_n) \nonumber \\
	& \pheq- \dot{a} \sum_{n=1}^{N+1} (n-1) (b_{n-1} \cos n\theta +
	c_{n-1} \sin n\theta)
	\nonumber \\ & \pheq
	+ a \sum_{m=1}^{N+1} (\dot{b}_{m-1} \cos m\theta
	+ \dot{c}_{m-1} \sin m\theta) \nonumber \\
	& \pheq 
	- \sum_{k=1}^N \sum_{m=0}^{N-k} \Big[(m+k)(b_{m+k}\dot{b}_m + c_{m+k}
	\dot{c}_m)\cos k\theta \nonumber \\
	& \pheq + (m+k) (c_{m+k} \dot{b}_m - b_{m+k} \dot{c}_m) \sin k\theta
	\nonumber \\ & \pheq
	+ m (b_m\dot{b}_{m+k} + c_m\dot{c}_{m+k}) \cos k\theta \nonumber \\
	& \pheq - m(c_m \dot{b}_{m+k} - b_m \dot{c}_{m+k})\sin k \theta \Big].
\end{align}
Equating the terms with different factors of sine and cosine yields
Eqs.~\eqref{eq:teva}, \eqref{eq:thighest}, \eqref{eq:tevsin}, and \eqref{eq:tevcos},
which together form the linear system that is used in the numerical integration
method.

~\\
~\\
~\\
~\\
\bibliography{adld}

\end{document}